\documentclass[12pt]{article} %***
\usepackage[sectionbib]{natbib}
\usepackage{array,epsfig,fancyheadings,rotating, amsmath, bm,algpseudocode, xr}
\usepackage[skip=0pt]{caption}

\usepackage[]{hyperref}  %<----modified by Ivan
\usepackage{sectsty, secdot}
\usepackage{lipsum}
\usepackage{blindtext}
\sectionfont{\fontsize{12}{14pt plus.8pt minus .6pt}\selectfont}
\renewcommand{\theequation}{\thesection\arabic{equation}}
\subsectionfont{\fontsize{12}{14pt plus.8pt minus .6pt}\selectfont}
\usepackage{amsmath}
\usepackage{amssymb}
\usepackage{amsfonts}
\usepackage{multirow}
\usepackage{amsthm}

\newcommand{\vv}[1]{\mbox{\boldmath $#1$}}
\theoremstyle{definition}

\usepackage{color}

\def\bsmal#1\nsmal{{\small\vspace*{-1mm}\begin{align}#1\end{align} }}

\makeatletter
\newcommand*{\addFileDependency}[1]{% argument=file name and extension
	\typeout{(#1)}
	\@addtofilelist{#1}
	\IfFileExists{#1}{}{\typeout{No file #1.}}
}
\makeatother

\newcommand*{\myexternaldocument}[1]{%
	\externaldocument{#1}%
	\addFileDependency{#1.tex}%
	\addFileDependency{#1.aux}%
}

\myexternaldocument{supplement_rev2_arxiv}
\begin{document}
	
	%%%%%%%%%%%%%%%%%%%%%%%%%%%%%%%%%%%%%%%%%%%%%%%%%%%%%%%%%%%%%%%%%%%%%%%%%%%%%%%%%%%%%%%%%%%%%%%%%%%%%%%%%%%%%%%%%%%%%%%%%%%%
	%%%%%%%%%%%%%%%%%%%%%%%%%%%%%%%%%%%%%%%%%%%%%%%%%%%%%%%%%%%%%%%%%%%%%%%%%%%%%%%%%%%%%%%%%%%%%%%%%%%%%%%%%%%%%%%%%%%%%%%%%%%%
	
	\renewcommand{\baselinestretch}{2}
	
%	\markright{ \hbox{\footnotesize\rm Statistica Sinica
			%{\footnotesize\bf 24} (201?), 000-000
	%	}\hfill\\[-13pt]
%		\hbox{\footnotesize\rm
%			%\href{http://dx.doi.org/10.5705/ss.20??.???}{doi:http://dx.doi.org/10.5705/ss.20??.???}
%		}\hfill }
	
	\markboth{\hfill{\footnotesize\rm Deborah Kunkel and Mario Peruggia} \hfill}
	{\hfill {\footnotesize\rm Anchored Bayesian mixture of regression models} \hfill}
	
	\renewcommand{\thefootnote}{}
	$\ $\par
	
	%%%%%%%%%%%%%%%%%%%%%%%%%%%%%%%%%%%%%%%%%%%%%%%%%%%%%%%%%%%%%%%%%%%%%%%%%%%%%%%%%%%%%%%%%%%%%%%%%%%%%%%%%%%%%%%%%%%%%%%%%%%%
	
	\fontsize{12}{14pt plus.8pt minus .6pt}\selectfont \vspace{0.8pc}
	\centerline{\large\bf Statistical inference with anchored Bayesian mixture \\}
	\vspace{2pt} 
	\centerline{\large\bf   of
		regressions models: An illustrative study of }
		\vspace{2pt} 
		\centerline{\large\bf 
			allometric data}
	\vspace{.4cm} 
	\centerline{Deborah Kunkel$^1$ and Mario Peruggia$^2$} 
	\vspace{.4cm} 
	\centerline{\it 	1. School of Mathematical \& Statistical Sciences, Clemson University, Clemson, SC, USA \\}
		\centerline{\it 
			2. Department of Statistics,
			The Ohio State University,
			Columbus, OH, USA\\}
	\vspace{.55cm} \fontsize{9}{11.5pt plus.8pt minus.6pt}\selectfont
	
	%%%%%%%%%%%%%%%%%%%%%%%%%%%%%%%%%%%%%%%%%%%%%%%%%%%%%%%%%%%%%%%%%%%%%%%%%%%%%%%%%%%%%%%%%%%%%%%%%%%%%%%%%%%%%%%%%%%%%%%%%%%%
	
	\begin{quotation}
		\noindent {\it Abstract:}
\\
	 We present an illustrative study in which we use a mixture of
	 regressions model to improve on an ill-fitting simple linear
	 regression model relating log brain mass to log body mass
	 for 100 placental mammalian species. The slope of the model
	 is of particular scientific interest because it corresponds
	 to a constant that governs a hypothesized allometric power
	 law relating brain mass to body mass. We model these data
	 using an anchored Bayesian mixture of regressions model,
	 which modifies the standard Bayesian Gaussian mixture by
	 pre-assigning small subsets of observations to given mixture
	 components with probability one. These observations (called
	 anchor points) break the relabeling invariance (or
	 label-switching) typical of exchangeable models. 
 In the article, we develop a
	 strategy for selecting anchor points using tools from case
	 influence diagnostics.  We compare the performance of
       three anchoring methods
 on the allometric data {and in simulated settings}.
%  and use auxiliary
%	 taxonomic information about the species to evaluate the
%	 model-based classifications estimated from
%	 these models.

		\vspace{9pt}
		\par
	\end{quotation}\par

	\def\thefigure{\arabic{figure}}
	\def\thetable{\arabic{table}}
	
	\renewcommand{\theequation}{\thesection.\arabic{equation}}

	\fontsize{12}{14pt plus.8pt minus .6pt}\selectfont

\section{Introduction}

In the natural sciences, allometry studies the relationships between
physical and physiological measurements taken on various animal
species \citep{peters1983ecological,Gayon2015}.  Of particular
interest is to determine how other measurements may be affected by
body mass.  Examples include the relationships between body mass and
brain mass, body mass and metabolic rate, body mass and gestation
duration.  It is often postulated that pairs $(x,y)$ of such
measurements may be related via a power law of the form $y = c x^b$,
for some unknown constants $c$ and $b$, typically assumed to be
positive.  The estimation of the exponent $b$ is often of primary
scientific interest.  On a logarithmic scale, the power law turns into
the linear relationship $\log y = (\log c) + b \log x$ and the
investigative focus shifts toward the estimation of the slope of the
regression line.
%These relationships have important implications on, for example,
%pharmacokinetics, 
%where allometric models can aid in understanding how metabolic rates
%change as children age\cite{}Wang2012}   

Given a set of $(x,y)$ pairs of traits
measured on a variety of species, it is by now
generally accepted that fitting a single linear regression model to
the entire data set provides too crude a summary, especially when many
species from different taxa and genetically diverse groups are
included in the data set
\citep{jerison1955brain,bennett1985relative,bennett1985brain}.  More
refined approaches rely on the incorporation of evolutionary
information (possibly inferred from a taxonomy) to perform an
analysis based on models for derived quantities that can be
treated as independent, rather than for the original measured traits
that exhibit species-related dependencies.  For example, this is the
case for a popular type of analysis based on
phylogenetically independent contrasts
\citep{felsenstein1985phylogenies,garland1992procedures}.
\cite{MacPer2002} show that traditional Bayesian variance components
models applied directly to allometric data for which taxonomic
information is available can produce a good fit and yield easily
interpretable inferences.  

In this article we reanalyze the data that \cite{MacPer2002} used to
illustrate their methods.  The data comprise the body and brain mass
measurements on 100 species of placental mammals originally reported
by \cite{sacher1974relation} as well as a taxonomy that assigns each
species to an order and sub-order based on its morphological and
physiological traits.  
%For example, the African lion (Panthera Leo) is
%listed as belonging to the order Carnivora and sub-order Fissipeda and
%the chimpanzee (Pan Troglodytes) is listed as belonging to order
%Primates and sub-order Anthropoidea.  
In total, the data contain
species that represent 13 orders and 19 sub-orders.

\begin{figure}
	\includegraphics[scale=.75]{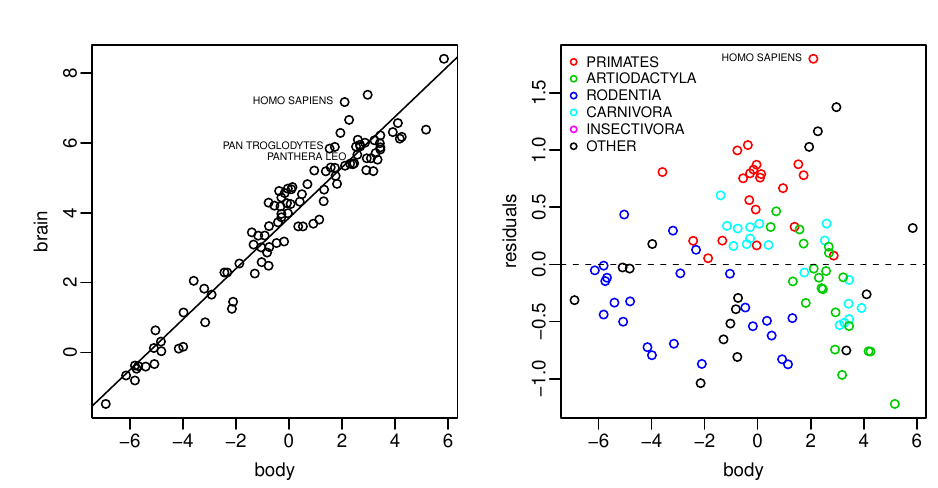}\caption{\label{figure:data_SLR}
		Mammals data (left) with the estimated least-squares
		regression line and residuals (right) from the least-squares
		regression fit.} 
\end{figure}

The data are shown in Figure~\ref{figure:data_SLR}. The left panel
displays a scatterplot of the centered log body
mass and log brain mass and the least-squares fit from a naive simple
linear regression 
model. The residuals for this model are shown in the right panel of
Figure~\ref{figure:data_SLR}.  This plot raises some concerns about
model fit.  First, there is a slight increase in residual variability
as log body mass grows.  Second, other features of the residuals can
be traced back to the species orders (distinguished by plotting
color): all Primates (red points) have positive residuals, while most
Rodentia (blue points) have negative residuals. This structure points
to the fact that the least squares line does not properly account for
within-order similarities in the allometric relationship.

\cite{MacPer2002} present a detailed evaluation that uncovers the
lack-of-fit of this model, introducing Bayesian model diagnostic
techniques based on case-deletion importance
sampling weights
\citep{geweke1989bayesian,bradlow1997case,peruggia1997variability,epifani2008case,thomas2018reconciling}.
They also present a Bayesian variance components model that includes
additive random effects for orders and
sub-orders. These random effects induce
positive correlations between the residuals of
species belonging to the same taxonomic groups and subgroups.
The random effect adjustment
effectively creates a separate regression line with its own
intercept for the species within each subgroup. \cite{MacPer2002}
show that this way of accounting for the
taxonomic information significantly ameliorates the quality of the
fit.

Suppose now that no information about the taxonomy were available. We
might wish to remedy a suspected lack-of-fit of the
simple linear regression model, but a random effects model is
not an option when the true group memberships are not known.  In such
situations, a reasonable modeling alternative is to assume that there
exist finitely many subgroups of observations for which separate
regression lines are appropriate, and yet it is unknown which
observations to associate with which regression.  This leads naturally
to the formulation of a mixture of regressions model in which,
%assuming the existence of a finite number of groups and 
conditional on {\em unobserved} group membership indicators,
observations falling in separate groups follow separate regression
models. 
 Using these data as the foundation for an illustrative study, we consider a Bayesian mixture of
regressions to account for unobserved heterogeneity in a data set.
The fundamental difference between the mixture setting and the random
effects setting considered by \cite{MacPer2002} is that, in the
former, group memberships are latent and are the subject of
inferential investigation.  An additional difference is that we will
allow for group-specific slopes in addition to group-specific
intercepts.

A unique challenge arises when modeling with Bayesian mixture models:
a typical Bayesian formulation would specify, a priori, a fully
exchangeable mixture, yielding a fully exchangeable posterior that is indifferent to any arbitrary relabeling of
the mixture components.
This phenomenon, often referred to as label switching, precludes
informative assignment of labels to the mixture components and prevents meaningful interpretation of the component-specific parameter estimates.  \cite{KunPer2020} introduce a modeling
device called anchoring that breaks the labeling invariance of the
exchangeable version of a Bayesian Gaussian mixture model.  The idea
is to identify small subsets of representative observations, called
the anchor points, and allocate them to separate components
before performing the analysis.  \cite{KunPer2020} characterize the
equivalence between this process and the specification of a weakly
data-dependent prior for the component specific parameters. A shrewd
selection of anchor points is essential for the successful
implementation of the strategy.  They propose a strategy for model
specification based on a modified EM algorithm.

Motivated by the mixture of regressions model and its application to
our allometric data, we extend the work of \cite{KunPer2020} in two
main directions.  First, we generalize the EM anchoring strategy to the case of mixture of
regressions models.  Second, we introduce a new strategy for the
selection of anchor points that builds on the work on case-deletion
analysis presented in \cite{MacPer2002} and
\cite{thomas2018reconciling}.  We compare this new strategy for the
selection of anchor points (which we call CDW-reg) to the EM
anchoring strategy based on the mixture of regressions model
(which we call EM-reg).
The strategies differ in the way they extract
information from the data to select the points
representative of the various mixture components.

In Section~\ref{section:model} we present the mixture of regressions
model and its corresponding anchor model.  In
Section~\ref{section:anchoredEM} we present the EM method for
selecting anchor points under the mixture of regressions model.  In
Section~\ref{section:acdw} we describe the CDW-reg method for choosing
anchor points via clustering of case-deletion weights.  The results of
our simulations and of the analysis of the mammals data are presented in
Sections~\ref{section:simulation} and~\ref{section:results}. A few final considerations are discussed in Section~\ref{section:discussion}.

\section{Background} \label{section:model}
%In this section we give a formal definition of the mixture of
%regressions model and its anchored version that we will employ to
%analyze the mammals data.
\subsection{Mixture of regressions model} \label{sec:mix-reg}
The $k$-component mixture of linear regressions model specifies that
the observations $(y_i, \vv{x}_i), i =1,\ldots,n$, are drawn from $k$
homogeneous subgroups within the population.  The response $y_i$ is a
scalar and $\vv{x}_i \in R^p$ is a row vector of predictors whose first
element is equal to one.  The data
are denoted by $\vv{y}^\prime = (y_1,\ldots,y_n)$ and
by the $n \times p$ matrix \vv{X} with $i$-th row equal to
$\vv{x}_i$. The mixture likelihood is
\begin{align}
\label{model_mixreg}
f( \vv{y} |\vv{X}, \vv{\beta}, \sigma^2, \vv{\eta}) &= \prod_{i=1}^n
\sum_{j=1}^k\eta_j  \phi(y_i;\vv{x}_i \vv{\beta}_j, \sigma^2),  
\end{align}
where 
$\vv{\eta}^\prime = (\eta_1, \ldots, \eta_k)$ is a vector of
mixture probabilities that satisfy
$\sum_{j=1}^k\eta_j=1$ and $\phi(\cdot; a,b)$ denotes the density
function of a normal distribution with mean $a$ and variance $b$,
evaluated at its argument. The component-specific regression parameters in this
model are $\vv{\beta} = (\vv{\beta}_1,\ldots,\vv{\beta}_k)$,
where $\vv{\beta}_j \in R^p$ is the vector of 
regression coefficients associated with the $j$th component.
This model can be written equivalently by introducing latent
allocations $\vv{s} = (s_1,\ldots,s_n)$, where $s_i=j$ indicates that
observation $i$ was generated from component $j$. Specifying that $P(S_i=j) = \eta_j$, for $j=1,\ldots,k$, produces the mixture of
regressions model in~\eqref{model_mixreg}. Conditional on
$s_i=j$, the mean of $y_i$
is $\vv{x}_i \vv{\beta}_j$ and, with the prior specification given below, the model is a random effects
regression.

Assuming conditional independence throughout, we specify the following exchangeable prior:
\begin{eqnarray} \label{prior} \vv{\beta}_j \;|\; \vv{\mu}_{\beta},
\vv{V} & \sim & N_p(\vv{\mu}_{\beta}, \vv{V}), \quad
j=1,\ldots,k,  \nonumber \\
\sigma^{-2} \;|\;a, b & \sim & \mbox{Gamma}(a,b), \\
\vv{\eta} & \sim & \mbox{Dirichlet}(\alpha \vv{1}_k), \nonumber
\end{eqnarray}  
where $\vv{V}$ is a $p \times p$ diagonal matrix whose $q-$th diagonal element is
$v_{q-1}$ and the Gamma distribution is parameterized to have mean $a/b$.
The exchangeable specification is a natural way of expressing
prior ignorance about the relation between $y$ and
$\vv{x}$ within each group: we make no assumptions that induce
different prior distributions on the $\vv{\beta}_j$. A
consequence of the exchangeable specification is that the posterior
density of $\vv{\beta}$ is invariant to relabeling: for any
permutation of the integers $1,\ldots,k$, denoted by $\rho_q(1:k)$,
$p(\vv{\beta}|\vv{y})$ is equal to
$p(\vv{\beta}_{\rho_q(1:k)}|\vv{y})$ for all~$\vv{\beta}$.
%In other words, the posterior density of a value of $\vv{\beta}$ is
%unchanged if the component labels are permuted. 
 This produces $k!$
symmetric regions in the posterior distribution of $\vv{\beta}$, each
corresponding to one of
the $k!$ possible labelings of the
mixture components. Inferentially, this is undesirable because marginal distributions of the component
specific parameters are identical, making it impossible to use estimates of the labeled parameters
$\vv{\beta}_1,\ldots,\vv{\beta}_k$ to infer
%distinctions
 differences
among
features of the $k$
groups.  
Computationally, the label switching
phenomenon hampers fitting the model by Markov chain Monte Carlo
simulation \citep{jasra2005markov}.

\subsection{Anchor models}
In previous work, \citep{KunPer2020} we have proposed a new class of
models called \textit{anchor models} which pre-classify some
observations in order to induce a non-exchangeable, data-dependent
prior which can alleviate the inferential nuisances caused by the model's posterior exchangeability.  
%To specify an anchor model, we select a small number of points to be
%assigned a priori to particular components of the 
%mixture model.  
The model is defined using $k$ index sets
$A_1,\ldots,A_k$, where $A_j$ contains the indices of a small number of observations
which will be be pre-labeled (or ``anchored'') to component $j$.  The number of
points in $A_j$, denoted by $m_j$, is chosen ahead of time and each
observation is anchored to at most one component; i.e.,
$A_j \cap A_{j^\prime} = \emptyset$ for $j \neq j^\prime$.  Given the
anchor points, $A = \cup_{j=1}^kA_j$, 
the anchored version of
model~\eqref{model_mixreg} arises by modifying the distribution on the
latent allocations, such that 
\vspace{-3mm}
\begin{eqnarray}
P(S_i=j) =\begin{cases}
1, \quad i \in A_j, \\[-8pt]
0, \quad i \in A_{j\prime}, j\prime \neq j, 
\end{cases}
\end{eqnarray}
\mbox{}
\vspace*{-9mm}
\mbox{}\\
and $P(S_i=j)=\eta_j$ for $i \not\in A$ as in the model~\eqref{model_mixreg}.
% has likelihood
%$f_A(\vv{y}|\vv{X}, \vv{\beta},\sigma^2,\vv{\eta} ) =
%\prod_{i=1}^nf_A(y_i|\, \vv{x}_i, \vv{\beta},\sigma^2,\vv{\eta})$,
%with 
%%\label{anchor_likelihood}
%\begin{eqnarray}
%\label{anchormodel_mixreg}
%%\begin{split}
%f_A(y_i|\vv{x}_i, \vv{\beta}, \sigma^2, \vv{\eta})  &=\begin{cases}
%\sum_{j=1}^k\eta_j\phi(y_i;\vv{x}_i\vv{\beta}_j; \sigma^2), \quad i \notin
%A, \\[-3pt] 
%\phi(y_i;\vv{x}_i \vv{\beta}_j; \sigma^2), \quad i \in A_j, \quad j=1,\ldots,k.
%\end{cases} 
%%\end{split}
%\end{eqnarray}
%Similarly, if the latent allocation variables are
%used, we have the equivalent representation
%$f_A(\vv{y}|\vv{X}, \vv{\beta},\sigma^2,\vv{\eta} ) =
%\prod_{i=1}^n\sum_{s_i=1}^kP_A(S_i = s_i)f(y_i| s_i, \vv{x}_i,
%\vv{\beta},\sigma^2,\vv{\eta})$, where
%$f(y_i|s_i, \vv{x}_i, \vv{\beta}, \sigma^2, \vv{\eta})
%=\phi(y_i;\vv{x}_i\vv{\beta}_{s_i}, \sigma^2)$, 
%$P_A(S_i = j ) = \eta_jI(i\notin A) + I(i\in A_j)$, $i=1,\ldots,n$,
%$j=1,\ldots,k$, and $I(\cdot)$ is an indicator function that equals 1
%when its argument is true and zero otherwise.

An anchored mixture model can be regarded as a hybrid between a random
effects model, in which the class membership is known for all
observations, and a pure mixture model, in which the class membership
is unknown for all observations.  The properties of this model depend
on which and how many anchor points are selected.  \cite{KunPer2020}
show that an anchor model can result in a nearly-unimodal posterior
density on the component-specific parameter when the anchor points are
judiciously chosen and at least $k-1$ components have anchor points.
%Further, they recommend selecting a small number of anchor points for each components, as large numbers of anchor points can lead to poor out-of-sample performance of the model. \textbf{possibly refine/fill in more backgr  ound here.}. 
They propose a modified expectation-maximization (EM) algorithm for
specifying anchor points for a multivariate Normal mixture model. The
next section describes an extension of this method to mixture of
regressions
models. % and is the basis for one of the models used in our study.

\section{Anchoring with the EM algorithm} \label{section:anchoredEM}
\cite{KunPer2020} propose the anchored
EM algorithm, a modified version of the EM algorithm for
maximum a posteriori estimation of anchored mixture models.
%For a generic vector
%$\vv{\gamma} = (\vv{\gamma}_1,\ldots,\vv{\gamma}_k)$ of , 
The standard EM algorithm for mixture models obtains estimates of
the model parameters $\vv{\theta}$ by iterating between an
E-step, which estimates the probability distribution $p(\vv{s})$ on
the latent allocations conditional on the current estimate of
$\vv{\theta}$, and an M-step, which updates $\vv{\theta}$ to maximize
the expected (with respect to $p(\vv{s})$) joint posterior density of
$\vv{\theta}$ and $\vv{s}$.  This approach can be viewed as
iteratively maximizing an objective function
$F(q,\vv{\theta}) = E_q\left(\log\left(
    p(\vv{y},\vv{\theta},\vv{s})/q(\vv{s})\right)\right)$ with respect
to a distribution $q(\vv{s})$ on the latent allocations (E step) and
$\vv{\theta}$ (M-step) \cite{Neal1998}.

The anchored EM algorithm
modifies the E-step by constraining the distribution $q(\vv{s})$
to correspond to an anchor model; that is, a distribution for which
$m_j$ observations are allocated to component~$j$ with
probability~1, for 
$j=1,\ldots,k$. 
%instead Instead, the E-step of the modified algorithm 
%estimate a
%probability distribution on
%$q(\vv{s})$ 
%that is restricted to corresponds to an anchor
%model for some choice of $A = \cup_{j=1}^kA_j$;
%that is, a distribution for which each
%$A_j$ specifies $m_j$ observations allocated to
%component~$j$ with probability~1.  
It can be shown that the optimal distribution $q(\cdot)$ is that which
is closest to the posterior distribution on $\vv{s}$ conditional on
$\vv{\theta}$ and $\vv{y}$ \citep{KunPer2020}. The resulting algorithm
recovers an anchoring structure that closely approximates the
unanchored posterior distribution of $\vv{\theta}$ near one of its
local modes.  

% parameter $\vv{\gamma}$ and
%mixture probability distribution
%$p(y_i|\vv{\gamma})=\sum_{j=1}^{k}\eta_jf(y_i|\vv{\gamma}_j)$.

\subsection{Anchored EM for the mixture of regressions}
The EM method 
for the mixture of regressions model~\eqref{model_mixreg} comprises
these steps.

%This algorithm is applied to a specific model by substituting the
%appropriate $j$-th component density for
%$f(y_i|\vv{\gamma}_j^{t-1})$ in
%Equation~\eqref{rij_definition} of the E-step
%above and deriving appropriate updates in the M-step.  The M-step updates the values of the component-specific parameters to
%maximize
%$\sum_{i=1}^n \sum_{j=1}^k \widetilde{r}^t_{ij}
%\log(f(\vv{y}|\vv{\gamma}, \vv{s})) + \log(p(\vv{\gamma})) +
%\log(p(\vv{\eta}))$, analogously to the standard EM algorithm for
%maximum a posteriori estimation.  In the next subsections we outline
%the steps of this algorithm for two mixture models that we will
%use in our analysis of the mammals data.
\begin{description}
	\item[Initialization.] Choose a small tolerance $>0$. Set
          $t=1$; $\Delta=100$. Initialize $\theta^0 = (\vv{\beta}^0,
          \sigma^{0}, \vv{\eta}^{0})$. \vskip -1mm
        \hspace*{-11mm}
	{\bf While $\Delta > \text{tolerance}$ do:}
	\item[E-step.] Calculate $r_{ij}^t$ for $i=1,\ldots,n$, $j=1,\ldots,k$, where $r_{ij}$ is the posterior probability that $S_i=j$ given $\vv{y},\vv{X},\vv{\beta}, \sigma, \vv{\eta} $, and satisfies
	\begin{eqnarray}
r_{ij}^t & \propto \eta_j^t \; \phi(y_i;\vv{x}_i \vv{\beta}_j^t, \sigma^{2t})
	\end{eqnarray} 
	\item[Anchor step.] For fixed values $m_j$, $j=1,\ldots,k$, update the anchor points by finding
	 $A^t = \cup_{j=1}^k A_j^t$ to maximize $\sum_{j=1}^k\sum_{i\in A_j}r_{ij}^{t},$ subject to $A_j \cap A_{j^\prime} = \emptyset$ and $|A_j|=m_j$ for all $j\neq j^\prime$.  Let
	 \begin{align}
	 \widetilde{r}^t_{ij} &= \begin{cases} r_{ij}^t \quad \quad &\text{ if } i \not\in A^t \\
	 1 \quad \quad&\text{ if } i \in A_j^t \\
	 0 \quad \quad &\text{ if } i \in A_{j^\prime}^t,\quad j^\prime \neq j 
	 \end{cases}
	 \end{align}
	\item[M-step.]  Update  $\theta^t = (\vv{\beta}^t, \sigma^{t}, \vv{\eta}^t)$ to maximize $F^*(q^t,\theta)$, where 
		\begin{eqnarray}  F^*(q^t,\theta) =  \log(p(\vv{\beta}, \sigma, \vv{\eta})) +  \sum_{j=1}^k  \sum_{i=1}^n\widetilde{r}^t_{ij}
	\log(\phi(y_i;\vv{x}_i \vv{\beta}_j, \sigma^{2})) , \end{eqnarray}
$p(\vv{\beta}, \sigma, \vv{\eta})$	is given in~\ref{prior}, and $F^*(q^t,\theta)$ is equal to $F(q^t,\theta)$ plus terms that are constant with respect to $\theta$. % $\argmax_{\theta} F(q^t,\theta) = \argmax_{\theta}F^*(q^t,\theta)$.
	\vskip 2mm 
	%Calculate $F(q^{t},\vv{\gamma}^{t})$ and 
	Update 
	$\Delta = F^*(q^{t},\theta^t)-F^*(q^{t-1},\theta^{t-1})$. Set $t = t+1$.
\end{description}

\noindent
{\bf End do}

\noindent Return $A_j^{t}, \, j=1,\ldots,k$ and $F^*(q^{t},\theta^{t})$.

\noindent Update steps and recommendations for initialization
and remediation of possible non-convergence
are
given in Supplement~1.  

%\textbf{One more paragraph about properties of EM?} 

\subsection{Anchoring on the mammals data} \label{subsection:mandk}
We now 
present an analysis of the mammals data using three different anchor
models, which differ in the method used to select the anchor
points. All models use the same number of mixture components ($k=3$)
and the same number of anchor points in each component
$(m=3)$. Selecting the number of mixture components is an open
research question, made more difficult by the intractability of the
mixture likelihood, poor theoretical properties of standard selection
criteria, and the non-identifiability of the model \citep{Roeder94,
  Nobile2004}. Several promising strategies that treat $k$ as a random
quantity have been developed in recent years \citep{MalWal2016,
  MilHar2018}, and these model-based approaches may be developed in
future work into methods that can specify both the anchor points and
the number of components.  In this study, the choice of three
components is motivated by a practical desire to restrain model
complexity while retaining sufficient flexibility to capture salient
morphological differences between groups of species. Traditional information indices support the 
choice of a small number of components: the BIC favors $k=2$ and increases as k ranges from 2 to 6 (BIC =192, 197, 211, 220, 233, for $k=2,\ldots,6$), while the AIC is smallest at $k=3$ with little evidence that $k\geq 4$ aids in model fit (AIC= 175.9, 173.6, 179.3, 180.7, 186.5 for $k=2,\ldots,6$).  Comments on a two-component fit are given in Section~\ref{section:discussion}. 

In selecting $m$, we confine the anchor points to be small subsets of
optimally chosen observations.  A minimum of one anchor point for
$k-1$ components is required to eliminate the prior exchangeability of
the model that leads to label-switching.  We exceed this minimum in
order to specify data-dependent prior information that can partially
characterize the locations and scales of underlying groups.  However,
we keep $m$ small because our previous work has demonstrated that when
the groups are not well-separated, using many anchor points can lead
to poor out-of-sample predictive performance \citep{KunPer2020}. The
sensitivity analysis outlined in
Supplement~3 indicates that inferences on
the mammals data are similar when we  reduce the number of anchor points from $m=3$ to $m=2$, but that
$m=1$ appears to provide insufficient prior information to identify
three distinct lines.

\subsection{EM anchoring on the mammals data} \label{section:aem_mammals}
\begin{figure}[t!]
	\includegraphics[scale=.85]{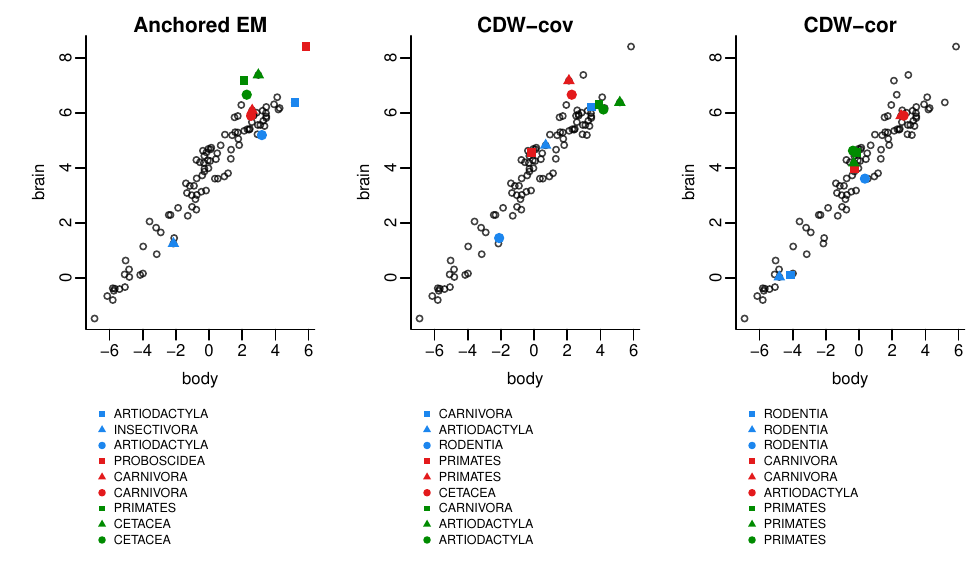}
	\caption{ \label{figure:selected_anchorpoints} Selected anchor
		points for the mammals data. The legend indicates the order of
		each anchor point.}
\end{figure}

We now apply the anchored EM method outlined in the previous
sections to the mammals data.  We set the prior hyperparameters
in~\eqref{prior} to be $a = 5$, $b = 1$,
$\vv{\mu}_\beta = (3.5, 0.6)^\prime$, and $v_0=1$, $v_1=0.5$. The left
panel of Figure~\ref{figure:selected_anchorpoints} shows the anchor
points selected. For ease of exposition, we code mixture grouping by color.
The anchored EM essentially performs an approximate maximum a
  posteriori fit of the mixture model, and it selects anchor points to
  be those with largest probability of allocation to their respective
  components.
Thus, anchor points assigned to the same component, are, in a sense,
those points that would be least likely to be assigned to either of
the remaining components.  The
anchor points for the red and blue groups exhibit some variability in
the x-direction and identify approximately parallel lines with
different intercepts.  The green points are clustered close together
in both the x- and y-directions, but suggest a line with a steeper
slope than that of the other two groups.  The auxiliary information
about the orders of the species selected as anchors, shown in the
bottom panel of Figure~\ref{figure:selected_anchorpoints}, indicates
that each component has two anchor points from a particular
order. This suggests that the anchor points may be partially
identifying the underlying structure that is driven by the taxonomic
information.

\section{Anchoring with case deletion weights} \label{section:acdw} 

In a Bayesian context, case-deletion analysis quantifies the influence
of an individual observation on the overall analysis by comparing the
posterior distribution conditional on the entire data set and the
posterior distribution conditional on the reduced data set obtained by
omitting the observation under consideration.  Case-deletion analysis
is an effective tool for identifying influential observations in
Bayesian models \citep{bradlow1997case,MacPer2002}.  It also provides
an avenue to assess the similarity of the impact of observations on
the inferential conclusions. We now derive a strategy for selecting
the anchor points of a Bayesian Gaussian mixture model as ``typical''
representatives of clusters identified via a preliminary
case-deletion analysis based on a model which assumes
a single regression line to describe all data points.

Consider a set of observations $\bm{y} = (y_1, \ldots, y_n)$ following
a model with density $f(\bm{y} | \bm{\theta})$ conditional on a set of
parameters $\bm{\theta}$ having prior distribution $\pi(\bm{\theta})$.
The posterior distribution of $\bm{\theta}$ given $\bm{y}$ is
proportional to the joint distribution of $\bm{y}$ and $\bm{\theta}$
which is
$p(\bm{y},\bm{\theta}) = f(\bm{y} | \bm{\theta}) \pi(\bm{\theta})$.
Similarly, denoting by $\bm{y}_{\setminus i}$ the reduced data set
obtained by deleting observation $i$, the posterior distribution of
$\bm{\theta}$ given $\bm{y}_{\setminus i}$ is proportional to
$p_{\setminus i}(\bm{y}_{\setminus i},\bm{\theta}) =
f(\bm{y}_{\setminus i} | \bm{\theta}) \pi(\bm{\theta})$.  

The ratio between the case-deleted and full posterior reacts to the
influence of the deleted case on the inferential conclusions.
For this reason it is useful to understand the behavior of the random
variables
\begin{equation}
w_i(\bm{\theta}) = \frac{p_{\setminus i}(\bm{y}_{\setminus
		i},\bm{\theta})}{p(\bm{y},\bm{\theta})}, \,\,\,\,\,\, i = 1,
\ldots, n, \label{eq:is_weights}
\end{equation}
when $\bm{\theta}$ follows the posterior distribution conditional of
the entire data set.
In 
practice, we can compute the normalized empirical versions of
the ratios in~(\ref{eq:is_weights}) using a sample
$\bm{\theta}_1, \ldots, \bm{\theta}_L$ 
from (approximately) the posterior distribution of $\bm{\theta}$
given $\bm{y}$. These quantities, known as case-deletion
importance sampling weights, are given by 
\begin{equation}
\bar{w}_i(\bm{\theta}_\ell) =
\frac{w_i(\bm{\theta}_\ell)}{\sum_{m=1}^Lw_i(\bm{\theta}_m)}, 
\,\,\,\,\,\, i = 1, \ldots, n; \,\,\, \ell = 1, \ldots, L.  
\label{eq:is_weights_norm}
\end{equation}  
%An important practical use of the normalized
%case-deletion weights is to reweight empirical averages based on
%samples from the full posterior to approximate expectations with
%respect to the case-deleted posterior \citep{geweke1989bayesian}.
%The behavior of the resulting weighted estimator is affected by
%the theoretical properties of the unnormalized weights
%in~(\ref{eq:is_weights}).  For example, a fundamental result in
%\cite{geweke1989bayesian} requires that the unnormalized weights
%have finite variance with respect to the full posterior for a
%central limit theorem for the weighted estimator to hold.

%In addition, 
The theoretical variability of the $w_i(\bm{\theta})$ and
the sample variability of the $w_i(\bm{\theta}_\ell)$ and
$\bar{w}_i(\bm{\theta}_\ell)$, $\ell = 1, \ldots, L$, are indicators of the
influence of observation $i$ on the posterior distribution, with
higher variability indicating larger influence
\citep{bradlow1997case,peruggia1997variability,epifani2008case}.  The
covariance matrix (with respect to the full posterior
distribution of $\theta$) of the log weights,
$\bm{C} = [C_{ij}] = [Cov(\log w_i(\bm{\theta}), \log
w_j(\bm{\theta}))]$, is a particularly useful quantity: $C_{ij}$ can
be interpreted as summarizing the degree of similarity between the
influence of deletion of case $i$ and case $j$.    

Further, for models of conditional independence,
\cite{thomas2018reconciling} detail the existence of a close
relationship between $\bm{C}$ and measures of influence based on
infinitesimal geometric perturbations of the multiplicative
contributions of each observation to the overall likelihood.  
In the
perturbed likelihood, the multiplicative factor corresponding to
each observation is raised to a power $\omega_i$, $i = 1 \ldots, n$.
The original likelihood is recovered by setting $\omega_i = 1$,
$i = 1 \ldots n.$ 
\cite{thomas2018reconciling} show that, in a
neighborhood of $\bm{\omega} = (1, \ldots, 1)^\prime$, $\bm{C}$
characterizes the curvature of the $n$-dimensional surface
representing the Kullback-Leibler divergence of the posterior based
on the original likelihood from the posterior based on the
geometrically perturbed likelihood.  Thus, $C$ contains information
about the directions in the $n-$dimensional real hyperplane along
which the Kullback-Leibler 
divergence surface changes more rapidly in response to geometric
likelihood perturbations.  This insight can be used to assess the
directional influence of cases.

As an exploratory tool for assessing such influence,
\cite{thomas2018reconciling} recommend to compute the sample
covariance matrix, $\widehat{\bm{C}}$, and the sample correlation
matrix, $\widehat{\bm{R}}$, based on the sample 
$\bm{\theta}_1, \ldots, \bm{\theta}_L$ and to perform a principal
component analysis (PCA) based on the eigendecompositions of these
matrices.  
% \add{.. As a result, we have a multivariate description of case
% influence }
The first several components are often sufficient to explain most of
the observed 
variability in the log weights and well-summarize the main directions
of influence.  A PCA display consisting of a scatterplot of the first
two or three normalized eigenvectors helps to reveal structure in the
data: points with high loadings in one or more components are
particularly influential, and points with similar loadings in all
components have similar influence. 

We leverage these ideas to form a strategy for choosing anchor points
based on the case-deletion weights.  First, we fit a single simple
linear regression model that does not accommodate latent heterogeneity
in the data.  
Graphical summaries based on a 
PCA eigendecomposition 
of $\widehat{\bm{C}}$ or
$\widehat{\bm{R}}$
can help 
to visualize clusters with similar directional
influence on the base model. To specify an appropriate mixture of
regressions model, we choose anchor points to be representatives of
these clusters. In this study we use k means
applied to the rows of 
$\widehat{\bm{C}}$ or
$\widehat{\bm{R}}$
to aid in identifying
clusters and representative points.

\subsection{CDW anchoring on the mammals data} \label{section:cdw_mammals}
\begin{figure}[h!]
	%\vspace{-0.7in}
	\includegraphics[width=.88\textwidth]{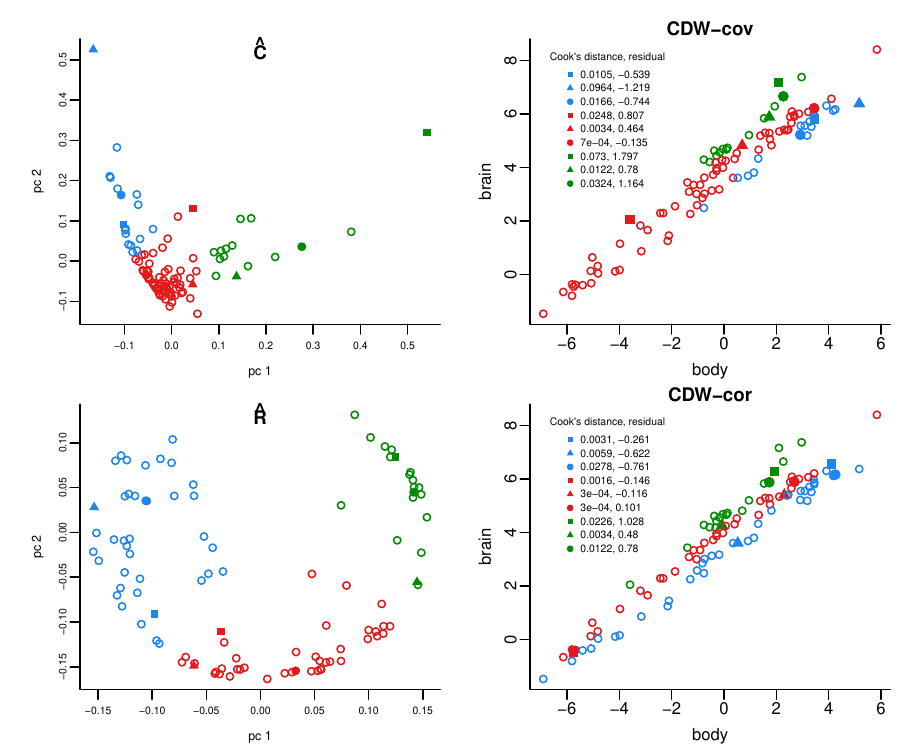}
	\caption{ PCA display based on the first two eigenvalues
		from the eigendecomposition of $\widehat{\bm{C}}$ (top) and $\widehat{\bm{R}}$ (bottom)
		of the log case-deletion weights. The resulting k
		means clustering with highlighted anchor points is shown in the right panel.} \label{fig:PCA}
\end{figure}

To apply the proposed technique to the analysis of the mammals data,
we began by fitting a Bayesian simple linear regression model with the
following hyperparameters: $a = 5$, $b = 1$,
$\vv{\mu}_\beta = (3.5, 0.6)^\prime$, and $v_0=1$, $v_1=0.5$.  A Gibbs
sampler was run to obtain 5,000 posterior samples after burn-in
and thinning of chains. The log case-deletion weights were computed
from the sampled residuals.  The left panels of Figure~\ref{fig:PCA}
shows the PCA displays based on the first two eigenvalues of the
eigendecomposition of $\widehat{\bm{C}}$ and $\widehat{\bm{R}}$.

Again postulating three mixture components, we ran the k means
clustering algorithm, %\citep{HarWon1979}
as implemented in R,
on the rows of 
$\widehat{\bm{C}}$ or
$\widehat{\bm{R}}$. Doing so,
we identified the red, green, and blue clusters
in the PCA display.  The scatterplots in the right panels of
Figure~\ref{fig:PCA} shows how the identified clusters map back into
observation space.  The $\widehat{\bm{C}}$-based clustering
appears to be reacting primarily to the size and sign of the
residuals from the fitted simple linear regression line, with the
green cases representing a small number of values tending to have
large, positive residuals and the blue cases representing a small
number of values with negative residuals. These green and blue cases
stand apart from most points in the PCA space, with the points with
largest residuals appearing well-separated from all other
points. The red cases, which comprise most of the data set, are
closely clustered in the PCA space and appear to be well
  described by the simple linear regression line.

  The $\widehat{\bm{R}}$-based PCA plot shows a nearly-circular shape
  and fewer points with high loadings in a particular direction. In
  the observation space, we see that the green and blue clusters
  correspond to points having rather large least squares residuals of
  the same sign within each cluster.

  To specify the CDW anchor models, we identified three representative
  observations within each cluster to be used as the three anchor
  points. To do this, for each of the initial clusters, we again ran
  the k means clustering algorithm to identify three sub-clusters.  A
  representative point from each sub-cluster was chosen as the point
  nearest to the sub-cluster centroid.  The selected points (three for
  each cluster) are shown as solid dots in the PCA displays and
  scatterplots of Figure~\ref{fig:PCA}.  In addition, the Cook's
  distances and residuals for the anchors points were obtained from the
  least-squares fit depicted in Figure~\ref{figure:data_SLR}. The
same nine points are also displayed in the middle and right panels of
Figure~\ref{figure:selected_anchorpoints}, along with the orders to
which the nine species belong. The model whose anchor points are
estimated from $\widehat{\bm{C}}$ is referred to as CDW-cov, and the
model whose anchor points are estimated from $\widehat{\bm{R}}$ is
referred to as CDW-cor.

In the CDW-cov model, 
%the points with highest residuals, such as the green square and the blue triangle, are set apart from the others and are selected as representatives to be anchor points. The additional anchor points for the blue and green clusters are less extreme outliers. They are separated, but all anchor points are close to each other in the x-y space.  
we see that, within each cluster, the set of anchor points comprises
some points with apparently large influence (the blue triangle
  and green square, for example, which have large Cook's distance
  values and large residuals, relative to the other points)
  and points that are less influential
  (such as the red triangle, whose x-value is near the sample average
  and which falls close to the regression line).  This feature
follows from the way in which k means decides to allocate member
species to the various sub-clusters, with some sub-clusters comprising
mostly ``usual'' observations and other, typically smaller,
sub-clusters comprising mostly ``unusual'' observations. Nonetheless,
the anchor points for the blue and green groups tend to be somewhat
close together in the x-y space; the larger residuals of points in
these clusters produces more distance in the PCA space.

In contrast, the CDW-cor anchor points do not stand out as
  unusual in the observation space and their Cook's distances tend to
  be smaller, with only two exceeding 0.02. These selected
points are widely separated in the x-direction, with each set of three
points hinting at a clearly distinct line (in a least squares
sense). These groups seem to 
be driven primarily by similar directions of influence.

\section{Simulation} \label{section:simulation} We performed a
simulation study to evaluate the performance of the anchored EM and
CDW methods in estimating parameters of three-component mixture of
regressions models. The details of the simulation design and its
results are presented in Supplement~2. In summary,
we found that the anchored EM models and CDW-cor models tend to
produce better accuracy in parameter estimation and cluster estimation
than CDW-cov. The weaker performance of CDW-cov can be explained by
the tendency for the points whose case-deletion weights have the highest variance to be separated from the others in the
$\widehat{\bm{C}}$ space. These
points, which are the most unusual
observations in  one or both of the x- and y- directions, are
often selected as anchor points by the  proposed k means method.
 However, they may not represent behavior
that would be typical of
any  other points,
leading to situations where some components describe
only the most unusual cases.
This phenomenon is alleviated when using the
CDW-cor method because the $\widehat{\bm{R}}$ matrix captures  a
  normalized version of the similarity of the influence of cases and
is less sensitive to the overall variability of the case deletion
weights. Future work may investigate alternatives to the use of the k
means clusters and sub-clusters to better select anchor points using
$\widehat{\bm{C}}$.

An interesting case considered in the simulation is one where data are
generated from three parallel lines (Setting B in
Supplement~2). This corresponds to a
random-intercepts regression when the source of heterogeneity is
unobserved. The CDW-cor method resulted in the most accurate
estimation in this case, suggesting that $\widehat{\bm{R}}$ can be a
useful tool in uncovering this type of latent heterogeneity. In
contrast, when the lines differ in slope and intercepts, anchored EM
performed better than the CDW methods.

% Reserve 3/4-1 page if possible. Results will be discussed briefly
% but design and most results will be in supplement.  \lipsum[1] Our
% results indicate that neither anchoring method is superior in all
% situations. When the components are well separated, the EM methods
% tend to before . In addition, the CDW method performs better when
% the correlation matrix of the case deletion weights is used than
% when the covariance matrix is used.

\section{Analysis of mammals} \label{section:results} 
%\subsection{Selection of anchor points}
%\subsection{AEM reg anchoring }
%\subsection{CDW-reg anchoring on the mammals data} 

%In addition to selecting the anchor points, the number of components and the number of anchor points per component must be selected. Choosing the number of components for a mixture model is an open area of research, and many strategies may be used. We selected $k=3$ components because \textbf{EXPLAIN.}
%\subsection{Results} \label{section:results} 
%The 
%\subsection{Parameter and cluster estimates}

We now present the inferential results obtained by fitting three
anchored mixture of regressions models to the mammals data: the
anchored EM (A-EM), CDW-cov, and CDW-cor models. We specify the prior
in~\eqref{prior} with the same hyperparameters specified in the
anchored EM selection method: $a=5$, $b=1$,
$\mu_\beta = (3.5, 0.6)^\prime$, and $v_0 = 1, v_1=0.5$.  For each
anchor model we ran a Gibbs sampler to obtain $L=7,500$ posterior
samples (after thinning and burn-in) of the model parameters and
assessed convergence using trace plots. To sample from an anchor model
with the Gibbs sampler, the latent allocation variables $s_i$ are
fixed for the anchor points and sampled only for the unanchored
points.

% Henceforth we refer to the specific models induced by the anchor
% points selected by the three selection methods as the A-EM, CDW-cov,
% and CDW-cor models.

\subsection{Parameter and cluster estimates} \label{section:estimates}
\begin{figure} 
	\includegraphics[width=.9\textwidth]{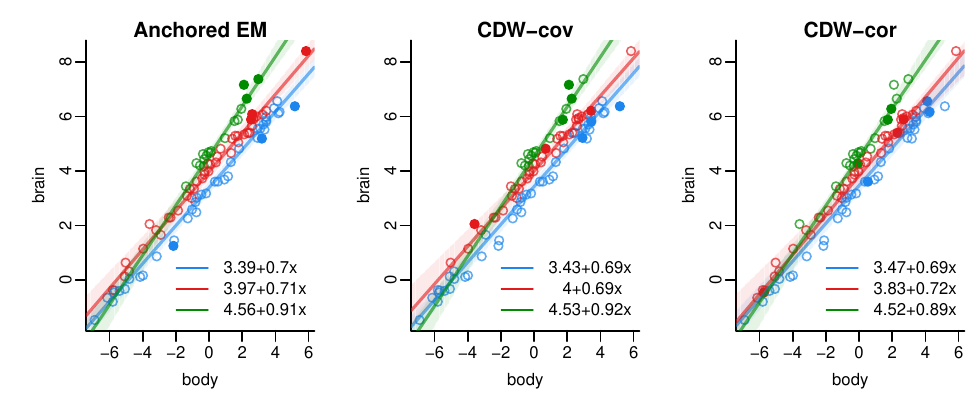}
	\caption{ \label{figure:estimated_regressions} Posterior mean
		regression lines and 90\% pointwise posterior credible intervals for the mean function for each mixture component. The data are color-coded
		according to their estimated allocation.}
\end{figure}

We used the posterior means of the model parameters to estimate
component regression lines and pointwise credible intervals for the component-specific mean functions, $\mathbf{X}\beta_j$, for each anchor model. 
%(These are the
%component-specific, posterior predictive regression lines.) 
These
lines are shown in Figure~\ref{figure:estimated_regressions}, with
Components~1,~2, and~3 drawn in blue, red, and green,  respectively.
We also estimated group membership for each species by finding a
maximum a posteriori estimate of its latent allocation, $s_i$.  Each data point in
Figure~\ref{figure:estimated_regressions} is
color-coded according to its estimated allocation, 
$\widehat{s}_i$.  The anchor points, whose group assignments
are assumed to be known, are shown as solid symbols to distinguish
them from the remaining observations.
% In Figure~\ref{figure:estimated_regressions} the estimated
% component regression lines based on the posterior means of the
% parameter estimates are shown for each set of anchor points.
% We also estimated a maximum a posteriori clustering structure in
% the data based on the posterior distribution of the latent
% allocations.  The cluster assignments,
% $\widehat{\vv{s}}=(\widehat{s}_1,\ldots,\widehat{s}_n)$, are
% calculated as $\widehat{s}_i = \max_{j}\widehat{p}(s_i = j|\vv{y})$
% where $\widehat{p}(s_i = j|\vv{y})$ is a Monte Carlo estimate of
% the posterior probability that observation $i$ is allocated to
% component~j.

All three models have identified a subgroup whose slope is
considerably steeper than those of the other groups, arbitrarily
labeled Component~3 (shown in green) for all models.  This group
contains species whose brain masses show a large increase
as body masses increase. The
estimated regression lines are similar across the methods, with
CDW-cor estimating the lowest slope (0.89) and CDW-cov estimating the
largest (0.92).  The three methods assign many of the same species to
Component~3, identifying a string of points in the upper-right end of
the scatterplot to be those arising from the steep
regression line. Fewer species are allocated to this group than to the
other two, and most are those with larger bodies from the Primate or
Cetacea orders.  In addition to these, the CDW-cor model assigns
several smaller species to Component~3, which appear in the
middle-left portion of the left panel of
Figure~\ref{figure:estimated_regressions} in an area  near the
  point where the estimated regression lines for Components~2 and~3
  intersect. These same points are assigned to Component~2 in the
CDW-cov model, and are split between Components~2 and~3 in the A-EM
model.

Component~1, plotted in blue in
Figure~\ref{figure:estimated_regressions}, represents, broadly
speaking, species with brains that are small relative to species of
the same size.  Of the 24 species in the Rodentia order, 18, 20, and
15 are allocated to this component by the A-EM, CDW-cov, and CDW-cor
models, respectively. For all models, this group has the smallest
intercept, and the estimated slopes are near 0.70 for all three anchor
models. Component~2, with regression lines drawn in red, has an
estimated slope nearly equal to that of Component~1. This group
represents the species which differ from the Component~1 species
mostly in their average brain size given their body size. The orders
of species allocated to Component~2 are varied, with Artiodactyla and
Carinorva being the most prevalent for all anchor models.

%The
%two other methods have assigned comparatively fewer observations to
%this component.  The EM-reg model estimates a Component~2 line with a
%larger intercept and shallower slope than the other methods. This
%difference can be attributed to the influence of the EM-reg anchor
%points, which fall in the center of the scatterplot; the green line
%fits these points quite well, as well as many other points in this
%region.  For the other two methods, both of which have anchor points
%with low $x$-values, the line fits the data well in the low-$x$
%range, and less well in the middle of the scatterplot.  In
%particular, there is a string of points corresponding to log body
%masses between $-1$ and 1 which all fall above the EM-MVN and CDW-reg
%lines. This suggests that additional mixture components might be
%needed to adequately describe the points not allocated to
%Components~1 and~2 under these models.

%\begin{figure}[t!]
%	\includegraphics[width=.9\textwidth]{clustering_byorder}
%	\caption{\label{figure:clusters_byorder} Data for the species
%		in orders Primates, Artiodactyla, and Rodentia color-coded
%		according to their model-based $\widehat{\vv{s}}$.  The rows
%		correspond to the three anchor models. }
%\end{figure}
\subsection{Validation and sensitivity} \label{section:validation}
For this case study, we specified a mixture model in order to
accommodate heterogeneity in the data due to taxonomic
differences. Because the auxiliary information about the species'
orders is available, we can evaluate the similarity in estimated
groups from the mixture model to the true orders of the species. In
Supplement~4, we demonstrate that all three
mixture models have estimated grouping that have some correspondence
with the known taxonomies: Group~1 tends to describe Rodentia, while
Group~3 seems to describe Primates and Cetacea.
\begin{figure}[h!]
	\includegraphics[scale=.9]{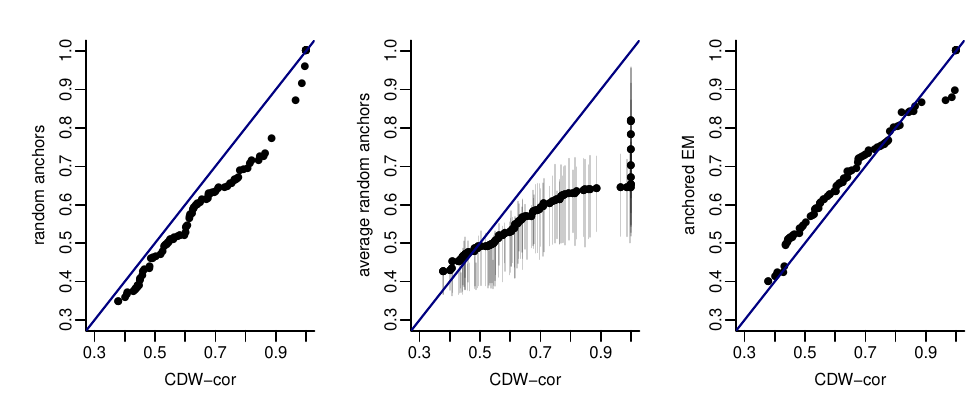}
		\caption{ \label{figure:maxprobs} Maximum allocation probabilities sorted by magnitude. The left panel compares the CDW-cor model to one randomly-selected anchor model. The middle panel displays the sorted maximum average allocation probabilities over 100 randomly-selected anchor models. The shaded lines show the 25th and 75th quantiles. The right panel compares the CDW-cor model to the A-EM model.}
\end{figure}

In the absence of auxiliary information, it may be desirable to
compare competing anchor models in terms of their effectiveness in
identifying distinct, well-separated mixture
components. \cite{KunPer2020} show that goodness-of-fit in an anchor
model is closely related to the degree of separation among the
component distributions: a well-fitting anchor model will produce
estimated mixture components with distinctive features. In contrast, a
poorly-fitting anchor model will exhibit features similar to the
``label-switching'' seen in full exchangeable models: densities of
component-specific parameters may be multimodal and similar in shape,
and estimated allocation probabilities will tend towards equal
probabilities for each component. Motivated by these considerations,
we propose to perform model-checking by summarizing the maximum
estimated posterior allocation probabilities calculated for each of
the observed data points; that is,
$p_i^{max}=\max_j \widehat{P}(S_i=j|y)$, $i=1,\ldots,n$.

Figure~\ref{figure:maxprobs} demonstrates three ways to assess the
model using the values of $p_i^{max}$. The right panel shows the order
statistics of the $p_i^{max}$ values from the CDW-cor model on the
x-axis and from the A-EM model on the y-axis. A straight line
would indicate that both models result in similar overall fit across
the two models. The fact that most points fall slightly above the identity line
indicates that most probabilities are higher for the A-EM model than for
CDW-cor, suggesting a better fit for the A-EM model.  If a competing
model is not considered, randomization can be used for checking one
anchor model. The left panel shows a plot that compares the
CDW-cor model to an anchor model whose anchor points are selected
completely at random. The CDW-cor probabilities are consistently
higher, resulting in a pattern with points falling underneath the
identity line. The middle panel is similar, but the y-axis shows the
average $p_i^{max}$ values over 100 randomized anchor models with shaded lines indicating the 25th and 75th sample percentiles of the $p_i^{max}$. This plot gives an approximation of the expected $p_i^{max}$ and their variability across random anchor models.

%While the AEM model results in classification probabilities that tend to exceed 1/3, favoring one component over the others, the classification probabilities for the random anchor points tend to be near 1/3, a feature that is similar to the full exchangeable model without resolution of label-switching. In practice, if the anchor model does not indicate, it may be desirable to use a different method for . Further, Figure~XX shows how the classification .

%We also calculated the values of $\widehat{\alpha}$, a measure proposed by \cite{KunPer2020} as a estimate of large-sample consistency of an anchor model. In small samples, this value similarly measures separation among components identified via the anchor points and can be used to compare competing models, with larger values indicating a high probability of estimated asymptotic posterior concentration on one permutation of the true model parameters.

A sensitivity analysis was performed to investigate the effect of the hyperparameters $a$, $b$, $v_0$, and $v_1$ in~\eqref{prior}. The analysis, which is detailed in Supplement~3, revealed parameter estimates for models with weaker prior information on $\sigma^{-2}$ and $\vv{\beta}$ that are similar to those in the analysis described in Section~\ref{section:estimates}. 

\section{Discussion} \label{section:discussion}
When we specify a finite mixture model, we assume the existence of
distinct subgroups in the population.  The group membership of any
individual observation is unknown and can be estimated a posteriori.
A random effects model similarly allows inference on group-specific
features, but requires auxiliary, deterministic information indicating
which observations are to be grouped together, and the similarities
among these observations drive the estimated features of the various
groups.  In an anchored mixture model, the anchored observations
affect the model fit in the same way as labeled observations in a
random effects model affect inferences on their groups: they inform
the distinct features of their mixture components.  These features
then influence the probabilities of group membership of the remaining
unanchored observations.  Thus, the similarities among a component's
anchor points are a key driver of what types of groups the mixture
model identifies.

In this case study, we selected three groups in order 
	to detect hypothesized subgroups tied to unobserved random effect indicators.
	A model with two mixture components could also have been specified to 
	to accommodate some of the unexplained structure in the residuals from the simple linear regression. Two components, while unable to capture many latent subgroups, can nonetheless distinguish between the most prominent inhomogeneous subsets of data points: for example, the primates such as homo sapiens are likely to be separated from the large-bodied, small-brained rodents, regardless of the number of components used.  The Supplement Section~3.1 gives details regarding the anchor points and estimated regression lines under a two-component model. A two component model may also be a first step in an iterative process, in which mixture components are added and model fit is re-evaluated to determine whether additional components are necessary. 

In our illustrative study, the EM method chose anchor points to fall
near the straight lines assumed by the mixture of regression
lines, and its accompanying mixture model defined groups based on
proximity 
to these lines.  For example, if we truly believe the mixture model we
have specified, a method such as anchored EM, which subsumes this
model at the anchoring stage, is perfectly appropriate. This method is
flexible and extensions to other regression models, including
non-Gaussian mixtures, can be readily derived within this framework.

The CDW methods are based on the principle that a mixture model is
appropriate because the simple model is inadequate; the mixture
components are identified as groups exhibiting similar misfit. In our
study, the CDW-cov method often selected points whose case deletion
weights have high variance. These points tend to have unusual
x-values and/or large residuals.  

 The CDW-cor method selected anchor
points that were representative of groups of observations exerting
similar influence on the posterior inferences from a naive model, and
our simulation results indicated that this method has promise in using
a mixture model to introduce random effects when auxiliary grouping
information is unobserved.  Further, because CDW requires only a
  base model and ability to calculate log case-deletion weights, this
  method can readily to be extended to non-Gaussian models, such as
  logistic or Poisson regression, and hierarchical models. Future
work will investigate how the theoretical properties of the
$\widehat{\bm{C}}$ and $\widehat{\bm{R}}$ matrices relate to the
optimal determination of the number of components and selection of
anchor points. 

In the multiple regression setting, we have demonstrated in simulations that the anchored mixture of regressions models can produce accurate parameter estimates in low dimensions, even if some predictors are  collinear. Further investigation is needed to evaluate how well the model scales when the number of predictors grows. In addition, real data sets are likely to exhibit features that affect the relative performance of the methods for selecting anchor points. For example, the current study illustrates that anchored-EM and the CDW methods differ in the variability among selected anchor points: CDW enforces some separation among the points, while anchored-EM may often select anchor points that are close together. We have found that in real data sets where some predictors may have low variability, discretized values, or inflation at zero, the differences among anchor models estimated from these methods may be magnified.

%
%The known taxonomic structure of species in our illustrative study provides
%insight about the model-based classifications produced by the three
%anchor models which extends beyond the particular data we analyzed.
%In our data, the groups of species from the same order exhibit
%distinct characteristics that might be primarily revealed by
%similarities in brain or body mass (the Artiodactyla species, for
%example, all have above-average body mass), by similarities in the
%change in brain mass as body mass changes, or both (as we see for the
%large primates).  The EM-reg model is particularly well-suited for identifying
%groups of the latter type because its anchor points are selected using
%the mixture of regressions, but it ignores differences in $x$ which
%provide valuable information for classification in this setting, where
%body mass is a distinguishing features of some of the taxonomic
%groups.  

In sum, to select a method for finding anchor points, it is
important to consider which types of distinguishing features the
underlying model uses to identify similar observations and to
determine how such features become relevant to answer the scientific
questions of interest.
	%%%%%%%%%%%%%%%%%%%%%%%%%%%%%%%%%%%%%%%%%%%%%%%%%%%%%%%%%%%%%%%%%%%%%%%%%%%%%%%%%%%%%%%%%%%%%%%%%%%%%%%%%%%%%%%%%%%%%%%%%%%%
%	\section*{Supplementary Materials}
%	
%	Contain
%	the brief description of the online supplementary materials.
%	\par
	%%%%%%%%%%%%%%%%%%%%%%%%%%%%%%%%%%%%%%%%%%%%%%%%%%%%%%%%%%%%%%%%%%%%%%%%%%%%%%%%%%%%%%%%%%%%%%%%%%%%%%%%%%%%%%%%%%%%%%%%%%%%
	\section*{Acknowledgements}
	
	This material is based on work supported by the National Science
	Foundation under grant no.~SES-1424481 and SES-1921523.
	\par
	
	%%%%%%%%%%%%%%%%%%%%%%%%%%%%%%%%%%%%%%%%%%%%%%%%%%%%%%%%%%%%%%%%%%%%%%%%%%%%%%%%%%%%%%%%%%

	\renewcommand{\baselinestretch}{1.5}

%\iffalse
\bibhang=1.7pc
\bibsep=2pt
\fontsize{9}{14pt plus.8pt minus .6pt}\selectfont
\renewcommand\bibname{\large \bf References}
%\begin{thebibliography}{11}
\expandafter\ifx\csname
natexlab\endcsname\relax\def\natexlab#1{#1}\fi
\expandafter\ifx\csname url\endcsname\relax
  \def\url#1{\texttt{#1}}\fi
\expandafter\ifx\csname urlprefix\endcsname\relax\def\urlprefix{URL}\fi
%\fi

	\renewcommand{\baselinestretch}{1}
\bibliographystyle{chicago}      % Chicago style, author-year citations
\bibliography{mybibfile_MP}   % name your BibTeX data base

%-------------------------------------------
%\vskip .65cm
%\noindent
%first author affiliation
%\vskip 2pt
%\noindent
%E-mail: dekunke@clemson.edu
%\vskip 2pt
%
%\noindent
%second author affiliation
%\vskip 2pt
%\noindent
%E-mail: (second author email)

\end{document}

	%%%%%%%%%%%%%%%%%%%%%%%%%%%%%%%%%%%%%%%%%%%%%%%%%%%%%%%%%%%%%%%%%%%%%%%%%%%%%%%%%%%%%%%%%%%%%%%%%%%%%%%%%%%%%%%%%%%%%%%%%%%%
	%%%%%%%%%%%%%%%%%%%%%%%%%%%%%%%%%%%%%%%%%%%%%%%%%%%%%%%%%%%%%%%%%%%%%%%%%%%%%%%%%%%%%%%%%%%%%%%%%%%%%%%%%%%%%%%%%%%%%%%%%%%%
	
	\renewcommand{\baselinestretch}{2}
	
	\markright{ \hbox{\footnotesize\rm Statistica Sinica
			%{\footnotesize\bf 24} (201?), 000-000
		}\hfill\\[-13pt]
		\hbox{\footnotesize\rm
			%\href{http://dx.doi.org/10.5705/ss.20??.???}{doi:http://dx.doi.org/10.5705/ss.20??.???}
		}\hfill }
	
	\markboth{\hfill{\footnotesize\rm Deborah Kunkel and Mario Peruggia} \hfill}
	{\hfill {\footnotesize\rm anchored Bayesian mixture of regression models} \hfill}
	
	\renewcommand{\thefootnote}{}
	$\ $\par
	
	%%%%%%%%%%%%%%%%%%%%%%%%%%%%%%%%%%%%%%%%%%%%%%%%%%%%%%%%%%%%%%%%%%%%%%%%%%%%%%%%%%%%%%%%%%%%%%%%%%%%%%%%%%%%%%%%%%%%%%%%%%%%
	
	\fontsize{12}{14pt plus.8pt minus .6pt}\selectfont \vspace{0.8pc}
	\centerline{\large\bf Supplement to Statistical inference with anchored Bayesian mixture \\}
	\vspace{2pt} 
	\centerline{\large\bf   of
		regressions models: An illustrative study of }
		\vspace{2pt} 
		\centerline{\large\bf 
			allometric data}
	\vspace{.4cm} 
	\centerline{Deborah Kunkel$^1$ and Mario Peruggia$^2$} 
	\vspace{.4cm} 
	\centerline{\it 	1. School of Mathematical \& Statistical Sciences, Clemson University, Clemson, SC, USA \\}
		\centerline{\it 
			2. Department of Statistics,
			The Ohio State University,
			Columbus, OH, USA\\}
	\vspace{.55cm} \fontsize{9}{11.5pt plus.8pt minus.6pt}\selectfont
	
	%%%%%%%%%%%%%%%%%%%%%%%%%%%%%%%%%%%%%%%%%%%%%%%%%%%%%%%%%%%%%%%%%%%%%%%%%%%%%%%%%%%%%%%%%%%%%%%%%%%%%%%%%%%%%%%%%%%%%%%%%%%%
%	
%	\begin{quotation}
%		\noindent {\it Abstract:}
%\\
%
%		
%		
%		\vspace{9pt}
%		\noindent {\it Key words and phrases:}
%Case-deletion weights, Clustering, EM algorithm
%		\par
%	\end{quotation}\par

	\def\thefigure{\arabic{figure}}
	\def\thetable{\arabic{table}}
	
	\renewcommand{\theequation}{\thesection.\arabic{equation}}

	\fontsize{12}{14pt plus.8pt minus .6pt}\selectfont

	\section{Anchored EM algorithm} \label{supplement:EM}

This section restates the steps of the anchored EM algorithm for the mixture of regressions model with  additional details on its implementation.
%\begin{description}
	\paragraph{Initialization.}
	To initialize $\theta^0 = (\vv{\beta}^0,
	\sigma^{0}, \vv{\eta}^{0})$, we recommend randomly partitioning the data into $k$ groups and initializing $\vv{\beta}$ at the least-squares estimates calculated from each group. The residual variance $\sigma^2$ can be initialized at the estimate from one of these least-squares solutions. We initialize $\Delta$ at a large positive value (100 in this study) and set the tolerance to a small, positive  value ($1\times 10^{-5}$ in this study).  \vskip -1mm
%	\hspace*{-11mm}
%	{\bf While $\Delta > \text{tolerance}$ do:}
	\paragraph{E-step.} Calculate $r_{ij}^t$ for $i=1,\ldots,n$, $j=1,\ldots,k$, where $r_{ij}$ is the posterior probability that $S_i=j$ given $\vv{y},\vv{X},\vv{\beta}, \sigma, \vv{\eta} $, and equals
	\begin{eqnarray} 
	r_{ij}^t & = \frac{\eta_j^t \; \phi\left(y_i;\vv{x}_i \vv{\beta}_j^t, \sigma^{2t}\right) }{\sum_{l=1}^k\eta_l^t \; \phi\left(y_i;\vv{x}_i \vv{\beta}_l^t, \sigma^{2t}\right)}.
	\end{eqnarray} 
where $	\phi(\cdot;a,b)$ denotes the density function of a normal distribution with mean $a$ and variance $b$. The distribution $q(\vv{s})$ is updated as
\begin{eqnarray}
q^t(\vv{s}) &=& \prod_{i=1}^nq^t(s_i) \\
&=& \prod_{i=1}^n\left(r_{ij}^t\right)^{I(s_i=j)}
\end{eqnarray}
where $I(s_i=j)$ is an indicator function that equals 1 if $s_i=j$ and equals 0 otherwise.
\paragraph{Anchor step.} For fixed values $m_j$, $j=1,\ldots,k$, update the anchor points by finding
	$A^t = \cup_{j=1}^k A_j^t$ to maximize 
	\begin{eqnarray}\label{anchor_step}\sum_{j=1}^k\sum_{i\in A_j}r_{ij}^{t},\end{eqnarray} subject to $A_j \cap A_{j^\prime} = \emptyset$ and $|A_j|=m_j$ for all $j\neq j^\prime$. 
Then set
	\begin{align}
	\widetilde{r}^t_{ij} &= \begin{cases} r_{ij}^t \quad \quad &\text{ if } i \not\in A^t \\
	1 \quad \quad&\text{ if } i \in A_j^t \\
	0 \quad \quad &\text{ if } i \in A_{j^\prime}^t,\quad j^\prime \neq j. 
	\end{cases}
	\end{align}
	The optimization step in \ref{anchor_step} simply amounts to assigning to
	component~$j$ the $m_j$ points with the highest posterior probability of
	allocation to component $j$ given the current estimate of the model
	parameters, in situations where this does not anchor any
	observations to more than one component. If an observation would be anchored to more than one component, a situation that could occur if the $m_j$ values are large relative to the sample size, an approximate solution may be used, or linear
	programming algorithms can produce an exact solution.

\paragraph{M-step.} In the M-step, we update $\theta^t = (\vv{\beta}^t, \sigma^{t}, \vv{\eta}^t)$ to maximize $F(q^t,\theta)$.
	The objective function satisfies
	\begin{eqnarray}
	F(q,\theta) & =  E_{q}\log(p(\vv{\theta},\vv{s},\vv{y} )) - E_q\log(q(\vv{s})),
	\end{eqnarray}
	with the second term constant with respect to $\theta$. The M-step is thus derived by maximizing the following with respect to $\theta$: 
	\small
	\begin{eqnarray}
 F^*(q^t,\theta)&= & E_{q}\log(p(\vv{\theta},\vv{s},\vv{y} ))  \\
	&=& E_{q}\log(f(\vv{y}|\vv{X},\vv{s},\vv{\theta})) + \log(p(\vv{\beta})+\log(p(\vv{\sigma^2}))+\log(p(\vv{\eta}))  \\	
	&=&   \sum_{j=1}^k  \sum_{i=1}^n\widetilde{r}^t_{ij}
	\log(\phi(y_i;\vv{x}_i \vv{\beta}_j, \sigma^{2})) + \log(p(\vv{\beta})+\log(p(\vv{\sigma^2}))+\log(p(\vv{\eta}))\\
	&= & \frac{n}{2}\log(\sigma^{-2}) -\frac{1}{2}\sum_{j=1}^k(\vv{y}-\bm{X}\vv{\beta}_j)^\prime \vv{R}_j^t(\vv{y}-\bm{X}\vv{\beta}_j)-\nonumber\\ &&\frac{1}{2}(\vv{\beta}-\vv{\mu}_{\beta})^\prime V^{-1}(\vv{\beta}-\vv{\mu}_{\beta}) + \nonumber\\ &&(a-1)\log\left( \sigma^{-2}\right) - b\sigma^{-2}+ (\alpha-1)\sum_{j=1}^k\log(\eta_j ) + c. \label{f_star}
	\end{eqnarray}
\normalsize In the expression above, $c$ represents terms that are constant with respect to $\theta$ and	$\vv{R}_j^t$ is an $n\times n$ diagonal matrix whose $i$-th diagonal
element is $\widetilde{r}_{ij}^t$.
	The update steps are as follows:
		\begin{eqnarray}
	\vv{\beta}_j^{t} & = & \left(\bm{X}^\prime \vv{R}_j^t\bm{X} + \vv{V}^{-1}\right)^{-1}\left(\bm{X}^\prime \vv{R}_j^t\vv{y} + \vv{V}^{-1}\vv{\mu}_{\beta}\right), \;\; j=1,\ldots,k, \nonumber \\
	\left(\sigma^{-2}\right)^t & = &  \frac{a+ n/2-1}{b + .5\sum_{j=1}^k\left(\vv{y}-\bm{X}\vv{\beta}_j^t\right)^\prime \vv{R}_j^t(\vv{y}-\bm{X}\vv{\beta}_j^t)}, \nonumber \\
	\eta_j^{t} & = & \frac{\sum_{i=1}^nr_{ij}^t + \alpha -1}{\sum_{l=1}^k\sum_{i=1}^nr_{il}^t + \alpha -1} , \quad j=1,\ldots,k. \label{etaj}
	\end{eqnarray} 
 
	%Calculate $F(q^{t},\vv{\gamma}^{t})$ and 

\paragraph{Monitoring convergence.} After the expectation, anchoring, and maximization steps, $t$ is increased by 1 and $\Delta$ is updated until its improvement does not exceed the tolerance.   To update 
	$\Delta$, we calculate $F^*(q^{t},\theta^t)-F^*(q^{t-1},\theta^{t-1})$. 	The function $F^*(q,\theta)$ is given by~\ref{f_star} above.
%		\begin{eqnarray}
%	F^*(q,\theta) & = & E_{q}\log(p(\vv{\theta},\vv{s},\vv{y} )) \nonumber \\
%	&=& E_{q}\log(f(\vv{y}|\vv{X},\vv{s},\vv{\theta})) + \log(p(\vv{\beta})+\log(p(\vv{\sigma^2}))+\log(p(\vv{\eta}))  \\
%	&\propto & \frac{n}{2}\log(\sigma^{-2}) -\frac{1}{2}\sum_{j=1}^k(\vv{y}-\bm{X}\vv{\beta}_j)^\prime \vv{R}_j(\vv{y}-\bm{X}\vv{\beta}_j)-\nonumber\\ &&\frac{1}{2}(\vv{\beta}-\vv{\mu}_{\beta})^\prime V_0^{-1}(\vv{\beta}-\vv{\mu}_{\beta}) + \nonumber\\ &&(a-1)\log\left( \sigma^{-2}\right) - b\sigma^{-2}+ (\alpha-1)\sum_{j=1}^k\log(\eta_j ).
%	\end{eqnarray}
%\end{description}
Convergence may instead be monitored using $F(q^{t},\theta^t)-F(q^{t-1},\theta^{t-1})$ by adding the term
$E_q\log(q^{t-1}(\vv{s})) - E_q\log(q^{t}(\vv{s}))$ to $\Delta$.

%\noindent Return $A_j^{t}, \, j=1,\ldots,k$ and $F(q^{t},\theta^{t})$.

\paragraph{Initialization and convergence.} 
We have found the algorithm to be sensitive to initial values and
prone to visit local maxima, especially when components are not
well-separated.  We thus recommend running the algorithm several times
(at least 15-20) and selecting the solution with the largest ending
value of $ F^*(q,\theta)$. We used 50 runs in the data analysis and
simulations in this study. For the 3-component mixture models
considered in this study, we have found that it is typical for about
half of the sequences to converge to the same solution.

In the analysis and simulations of this manuscript, we set the
tolerance for the algorithm to be 1$\times 10^{-5}$ and typically saw
convergence in less than 200 iterations. If the algorithm has not
converged after a very large number of iterations, it should
be stopped and
re-started from new random starting points.  We have found that,
more often, the algorithm may converge in 3 or 4 iterations to a
``poor'' solution, such as one in which two components have identical
parameters. If this behavior is observed in the final solution, the
algorithm should be re-started from new random starting points;
however, we have found that this situation is prevented by
using several runs 
and retaining only the best solution.

\section{Simulation study} \label{supplement:sim}
	
We conducted a simulation study to evaluate the performance of the
three anchoring methods: A-EM, CDW-cov, and CDW-cor.
\subsection{Data generation.}
Data sets were generated under mixtures of simple linear regressions
models with $k=3$ components.
To mimic the mammals data studied in the main text, we used sample
sizes of $n=100$ for all settings. Six settings were used to generate
the data: A1, A2, B1, B2, C1, and C2. Settings A, B, and C used
different values of $\vv{\beta}$, inducing different relationships
among the true regression lines. Figures~\ref{modelsA},~\ref{modelsB},
and~\ref{modelsC} show the true regression lines for each case.  One
additional setting with a multiple regression model is considered in
Section~\ref{section:simD}.
	
For each value of $\vv{\beta}$, we used two values of $\sigma$. These values were chosen so that the ratio 
	\begin{align}
	\frac{\sum_{j=1}^k\eta_j(\vv{\beta}_j-E_s(\vv{\beta}))^2}{\sum_{j=1}^k\eta_j(\vv{\beta}_j-E_s(\vv{\beta}))^2 + \sigma^2},
	\end{align}
which reflects the ratio of expected across-cluster
variability to total variability, was set to be 0.95 (settings A1, B1,
and C1) and 0.8 (settings A2, B2, and C2).
	
For each setting, 100 data sets were generated at random. The
predictor variable, $x$, was simulated from a standard normal
distribution and centered and rescaled. Latent allocations were
sampled with $\eta_1=\eta_2=\eta_3=1/3$, and data were sampled
conditional on the latent allocations.
        
As we did in the analysis of the mammals data performed in the main
text, $m = 3$ anchor points per component were selected using the
A-EM, CDW-cov, and CDW-cor methods. Each anchor model was fit using a
Gibbs sampler and posterior means of the model parameters were
estimated. The hyperparameters were fixed at the following values:
$a=5$, $b=1$, $v_0=1$, $v_1=3$, $\alpha=1.5$,
$\vv{\mu}_{\beta}=(\bar{y},0)^{\prime}$, where $\bar{y}$ is the sample
mean of the simulated response variable.  After fitting the models,
posterior means were computed for the model parameters, and maximum a
posteriori estimates were obtained for the latent allocation.

 In addition to the three anchor models, we analyzed each data set
 using a traditional mixture of regressions model with no anchor
 points. Post-hoc relabeling was used to relabel the posterior
 samples. This procedure involves first fitting a mixture of
 regressions model with no anchor points and then applying the
 likelihood-based relabeling method strategy of \cite{stephens2000} as
 implemented in the R package \texttt{label.switching}
 \citep{R_labelswitching}. The package does not automatically relabel
 the sampled allocation vectors, so estimated allocations were not
 computed for this method.
	
 \paragraph{CDW implementation details.} When implementing the CDW
 methods, anchor points were chosen automatically using k means on
 the rows of the $\widehat{\bm{C}}$ and $\widehat{\bm{R}}$
 matrices, and then k means was run a second time to find
 sub-clusters and their centroids. In order to automate the procedure,
 we did not make PCA displays. We recommend including this step in
 analysis of real data, however, to evaluate graphically the
   features of the selected anchor points.
	
   For some simulated data sets, the CDW method occasionally estimated
   initial clusters with too few observations to select 3 anchor
   points per component. This was more typical of the CDW-cov method,
   due to k means identifying a small cluster of points with large
   variances of the log case deletion weights.  In these situations,
   we still selected three anchor points per component by artificially
   adding points to the too-small cluster. The points that were
   closest to the centroid in Euclidean distance were artificially
   added until each cluster contained at least 5 points. This decision
   was made to facilitate the automation of the procedures. In
   practice, if this occurs, it may be appropriate to instead modify
   the model to require fewer anchor points for the affected
   component(s).  \vskip 3mm

\subsection*{Evaluation}
We evaluated the methods' performance on the simulated data sets by
measuring estimation accuracy and clustering accuracy.
\paragraph{Squared estimation error.}  Monte Carlo estimates of the
posterior means were used as the primary estimates of each model
parameter. We denote by $\widehat{\theta}_d$  the estimated posterior
mean of a parameter $\theta$ from simulated data set $d$.

To calculate error, we first relabeled the posterior means to minimize
the error in estimating $\vv{\beta}$. Relabeling is not typically
necessary in an anchor model; however, this step ensured that it was
possible to compare estimated parameters to the true values that
generated the data. For each possible relabeling of the
component-specific parameters, the error was calculated as the sum of
the relative squared distances,
$\left((\widehat{\theta}_d-\theta_d)/\theta_d\right)^2$, of the
posterior means from their true values. The relabeling that minimizes
this value was chosen as the final parameter estimate.

After achieving the optimal relabeling of parameter estimates, we
calculated the mean squared error separately for the intercept
parameters, slope parameters, mixture weights, and residual
variance. The mean squared error for vector-valued $\vv{\theta}_d$ was
calculated
as
$$\frac{1}{D}\sum_{d=1}^D(\vv{\theta}_d-\vv{\theta})^T(\vv{\theta}_d-\vv{\theta}). $$
%	\begin{description}
%		\item[Error in $\sigma^2$:] $\frac{1}{D}\sum_{d=1}^D(\widehat{\sigma}_d^2-\sigma^2)^2.$
%		\item[Error in $\beta_0$.] $\frac{1}{D}\sum_{d=1}^D\sum_{j=1}^k\left(\widehat{\beta}_{0jd}-\beta_{0j}\right)^2 $
%		\item[Error in $\beta_1$.] $\frac{1}{D}\sum_{d=1}^D\sum_{j=1}^k\left(\widehat{\beta}_{1jd}-\beta_{1j}\right)^2 $
%		\item[Error in $\eta$.] $\frac{1}{D}\sum_{d=1}^D\sum_{j=1}^k\left(\widehat{\eta}_{jd}-\eta_{sj}\right)^2 $
%	%	\item[Total error.]
%%		 $\frac{1}{D}\sum_{d=1}^D\left((\widehat{\sigma}_d^2-\sigma^2)^2+
%%		\sum_{j=1}^k\left( \left(\widehat{\beta}_{0jd}-\beta_{0j}\right)^2+\left(\widehat{\beta}_{1jd}-\beta_{1j}\right)^2+ \left(\widehat{\eta}_{jd}-\eta_{sj}\right)^2 \right) \right) $ 
%	\end{description}
\paragraph{Clustering accuracy.} We also assessed the accuracy of the maximum a~posteriori estimates of $\vv{s}$ from each anchor model. To measure accuracy, we calculated the Rand index between the estimated allocation to the true allocation that generated the data. The Rand index measures the similarity of two clustering structures estimated from the same data \citep{rand}.
Numbers close to 1 indicate that the anchor model's clustering is similar to the allocation that in truth generated the data. Numbers close to 0 indicate a dissimilar grouping of the observations. 

As a benchmark, we also estimated an allocation using likelihood-based classification probabilities evaluated at the true values of the model parameters:  
\begin{align}
\label{oracle_allocation}
s^{oracle}_i = \max_{j} P(S_i=j|y_i,\vv{\beta},\vv{\eta},\sigma^2) , \quad i=1,\ldots,n
\end{align}
where 
\begin{align}
P(s_i=j|y_i,\vv{\beta},\vv{\eta},\sigma^2)&\propto   \eta_j\phi(y_i; x_i\beta_j,\sigma^2).
\end{align} We refer to the allocation vector estimated using \ref{oracle_allocation} as the ``oracle'' allocation.

\subsection{Results: Setting A}

Figure~\ref{modelsA} shows the regression lines used to simulated data
in settings~A1 and A2 (left panel). The center and right panels show
examples of simulated data sets from these two settings. 

\vskip 3mm
%\FloatBarrier
%\begin{figure}
\includegraphics[width=.98\textwidth]{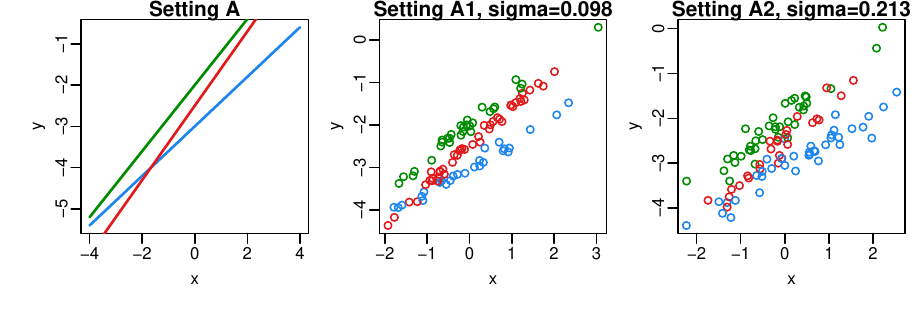}
\captionof{figure}{\label{modelsA}True regression lines and
  representative simulated data sets for Setting~A. Plotting symbols
  are colored by their true allocation.}
%		\end{figure}
%
%\FloatBarrier
\hspace*{1.7pc} The plots in Figure~\ref{fig:A1summary} summarize the
clustering and estimation performance of the four methods for setting
A1. The left panel shows boxplots of the Rand index for the three
anchor models and the oracle allocation. Among the anchor models, the
CDW-cor method has the highest median clustering accuracy and, for a
few data sets, achieves values close to those typical of the oracle
allocation. The A-EM and CDW-cov models perform similarly to each
other, with more variability in the EM values.

	\includegraphics[width=.98\textwidth]{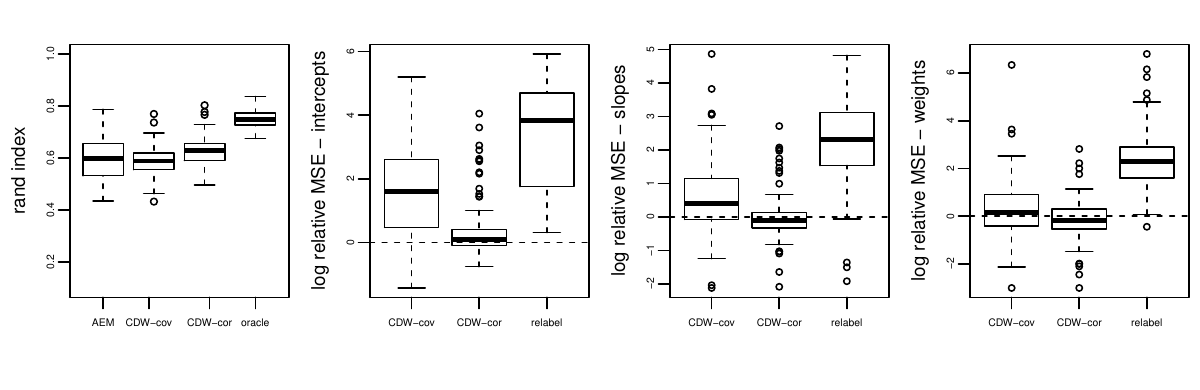}
	\captionof{figure}{\label{fig:A1summary} Results from setting A1: Rand index (left) and MSE relative to anchored EM for the intercept parameters (left center), slope parameters (right center), and mixture weights (right). }

        \hspace*{1.7pc} The right panels of Figure~\ref{fig:A1summary}
        show boxplots of the log MSE relative to the A-EM model under
        Setting~A1. Values greater than zero indicate poorer
        performance than the A-EM model and values less than zero
        indicate better performance. The left center plot shows that
        the CDW-cov and relabeling method typically have much higher
        error than A-EM, while CDW-cor outperforms A-EM about 25\% of
        the time. In estimating the slope (right center panel),
        CDW-cor performs comparably to A-EM, with CDW-cov and
        relabeling again resulting in larger error. All of the anchor
        models show similar performance in estimating the mixture
        weights, $\vv{\eta}$.

	\includegraphics[width=.98\textwidth]{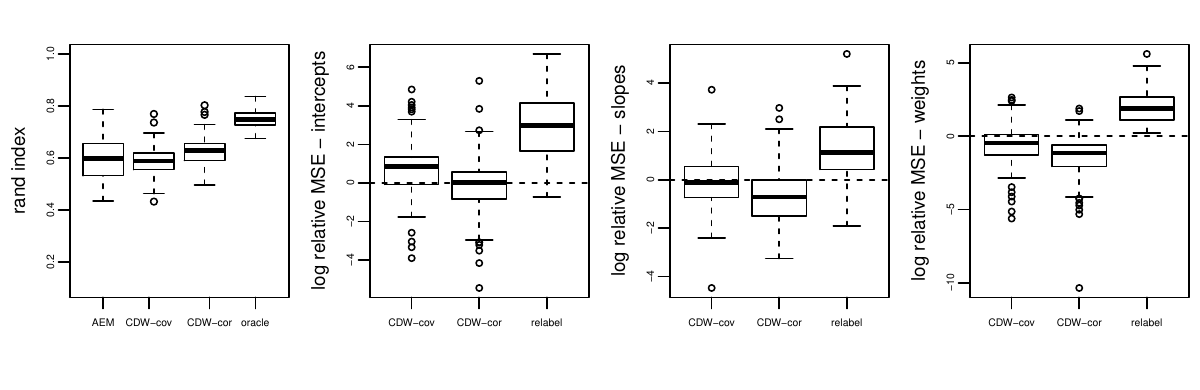}
	\captionof{figure}{\label{fig:A2summary} Results from setting A2: Rand index (left) and MSE relative to anchored EM for the intercept parameters (left center), slope parameters (right center), and mixture weights (right). }

        \hspace*{1.7pc} Figure~\ref{fig:A2summary} shows the same
        summaries for setting A2, where the residual variability is
        higher. In this more difficult case, the methods perform
        similarly in classification accuracy, with CDW-cor again
        tending to be the best. In this setting, CDW-cor improves in
        estimation accuracy relative to A-EM, and in fact out-performs
        A-EM when estimating the slopes. The CDW-cov also exhibits
        better relative performance, but still has poorer estimation
        accuracy than A-EM. As in the previous setting, the relabeling
        method provides the least accuracy.

        \hspace*{1.7pc} The average posterior means of each parameter
        are shown in Table~\ref{sA:estimates} for settings A1 and
        A2. These values indicate that the relabeling method tends to
        estimate one component with very low weight. This component is
        estimated to have a shallow slope and intercept between that
        of the other two components. An explanation for this is that
        the method identifies some red points between the green
          and blue lines, as seen in Figure~\ref{modelsA}, as
        belonging to a distinct component, but fails to accurately
        detect the linear pattern followed by the red points
        across a wider range of x-values 
        without prior information from the anchor points.  \vskip 4mm
        \captionof{table}{ \label{sA:estimates}Average posterior means
          for each model under setting A1 (top) and A2
          (bottom). Values are the posterior means for each parameter,
          averaged over all data sets. Values in parentheses are
          estimated standard errors.}
	\begin{adjustbox}{width=\textwidth}
		\begin{tabular}{rllllllllll}
						\hline
			\hline
						\multicolumn{1}{l}{}&		\multicolumn{10}{l}{\textit{Setting A1}}\\

			\hline
					\multicolumn{1}{l}{}&		\multicolumn{3}{|c|}{Component 1}&\multicolumn{3}{|c|}{Component 2}&\multicolumn{3}{|c|}{Component 3}&\multicolumn{1}{c}{}\\
			\hline
	& $\beta_{0}$ & $\beta_{1}$ & $\eta$ &  $\beta_{0}$ & $\beta_{1}$ & $\eta$ &  $\beta_{0}$ & $\beta_{1}$ & $\eta$ & $\sigma^2$ \\ 
			\hline
true &    -3 &   0.6 & 0.3333 &  -2.5 &   0.9 & 0.3333 &    -2 &   0.8 & 0.3333 & 0.009591 \\ 
AEM & -2.948 (0.05)  & 0.577 (0.04)  & 0.343 (0.05)  & -2.485 (0.06)  & 0.932 (0.07)  & 0.321 (0.07)  & -2.062 (0.03)  & 0.769 (0.03)  & 0.335 (0.06)  & 0.049 (0.01)  \\ 
CDW-cov & -2.862 (0.14)  & 0.613 (0.13)  & 0.323 (0.06)  & -2.553 (0.23)  & 0.807 (0.16)  & 0.304 (0.05)  & -2.152 (0.10)  & 0.808 (0.07)  & 0.373 (0.07)  & 0.064 (0.01)  \\ 
CDW-cor & -2.936 (0.08)  & 0.589 (0.06)  & 0.338 (0.05)  & -2.494 (0.10)  & 0.905 (0.10)  & 0.311 (0.06)  & -2.083 (0.07)  & 0.770 (0.06)  & 0.351 (0.06)  & 0.051 (0.01)  \\ 
relabel & -2.910 (0.04)  & 0.566 (0.03)  & 0.377 (0.06)  & -2.460 (0.04)  & 0.601 (0.17)  & 0.168 (0.13)  & -2.195 (0.09)  & 0.785 (0.10)  & 0.455 (0.12)  & 0.069 (0.01)  \\ 
\hline
			\hline
						\hline
			\multicolumn{1}{l}{}&		\multicolumn{10}{l}{\textit{Setting A2}}\\
			\hline
						\hline
			\multicolumn{1}{l}{}&		\multicolumn{3}{|c|}{Component 1}&\multicolumn{3}{|c|}{Component 2}&\multicolumn{3}{|c|}{Component 3}&\multicolumn{1}{c}{}\\
			\hline
			& $\beta_{0}$ & $\beta_{1}$ & $\eta$ &  $\beta_{0}$ & $\beta_{1}$ & $\eta$ &  $\beta_{0}$ & $\beta_{1}$ & $\eta$ & $\sigma^2$ \\ 
			true &    -3 &   0.6 & 0.3333 &  -2.5 &   0.9 & 0.3333 &    -2 &   0.8 & 0.3333 & 0.04556 \\ 
			AEM & -2.920 (0.07)  & 0.558 (0.08)  & 0.318 (0.08)  & -2.479 (0.15)  & 0.874 (0.20)  & 0.350 (0.09)  & -2.096 (0.10)  & 0.765 (0.11)  & 0.332 (0.11)  & 0.095 (0.01)  \\ 
			CDW-cov & -2.842 (0.16)  & 0.602 (0.14)  & 0.309 (0.07)  & -2.541 (0.20)  & 0.783 (0.20)  & 0.320 (0.07)  & -2.182 (0.15)  & 0.810 (0.10)  & 0.371 (0.08)  & 0.106 (0.02)  \\ 
			CDW-cor & -2.917 (0.11)  & 0.608 (0.08)  & 0.329 (0.05)  & -2.467 (0.14)  & 0.828 (0.14)  & 0.324 (0.05)  & -2.118 (0.09)  & 0.797 (0.07)  & 0.347 (0.05)  & 0.095 (0.01)  \\ 
			relabel & -2.841 (0.09)  & 0.569 (0.06)  & 0.361 (0.08)  & -2.468 (0.05)  & 0.491 (0.14)  & 0.113 (0.10)  & -2.256 (0.08)  & 0.830 (0.10)  & 0.526 (0.12)  & 0.125 (0.02)  \\ 
				\hline
		\end{tabular}

	\end{adjustbox}

\clearpage
\subsection{Results: Setting  B}
Settings B1 and B2 generated data from three parallel lines, as
shown in Figure~\ref{modelsB}.  \vskip 4mm
\includegraphics{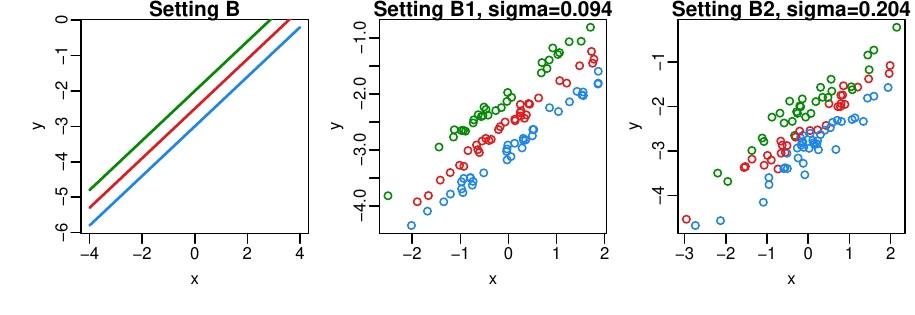}
\captionof{figure}{\label{modelsB} True regression lines and
    representative simulated data sets for Setting~B. Plotting
  symbols are colored by their true allocation.}

\hspace*{1.7pc} The performance summaries under setting B1 are shown
in Figure~\ref{fig:B1summary}. The typical values of the Rand index
are highest for the CDW-cor models, and the MSE is lowest for the same
method. Figure~\ref{fig:B2summary} shows that in Setting B2, the case
with less separation, the advantage of CDW-cor in terms of MSE is even
stronger. CDW-cov in this setting performs as well or better than
anchored EM, in contrast to settings where models have differing
slopes. 

\vskip
4mm \includegraphics[width=.98\textwidth]{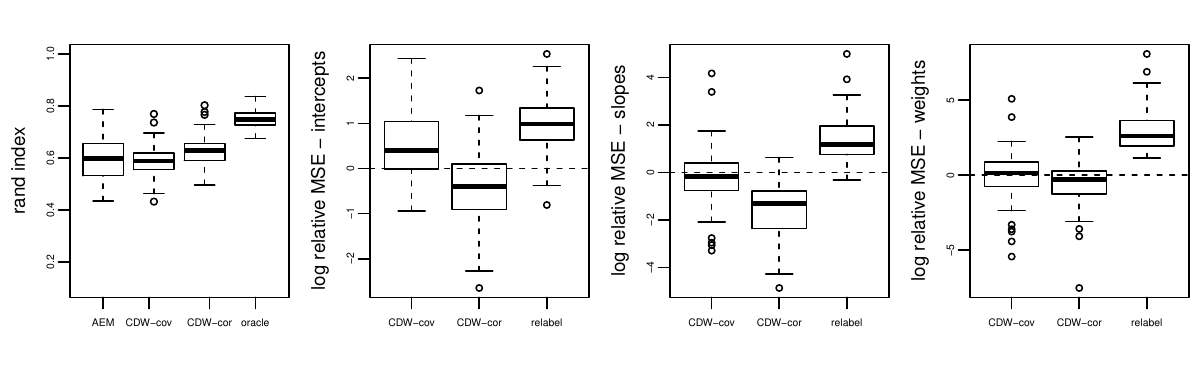}
\captionof{figure}{\label{fig:B1summary} Results from setting B1: Rand
  index (left) and MSE relative to anchored EM for the intercept
  parameters (left center), slope parameters (right center), and
  mixture weights (right). }

	\includegraphics[width=.98\textwidth]{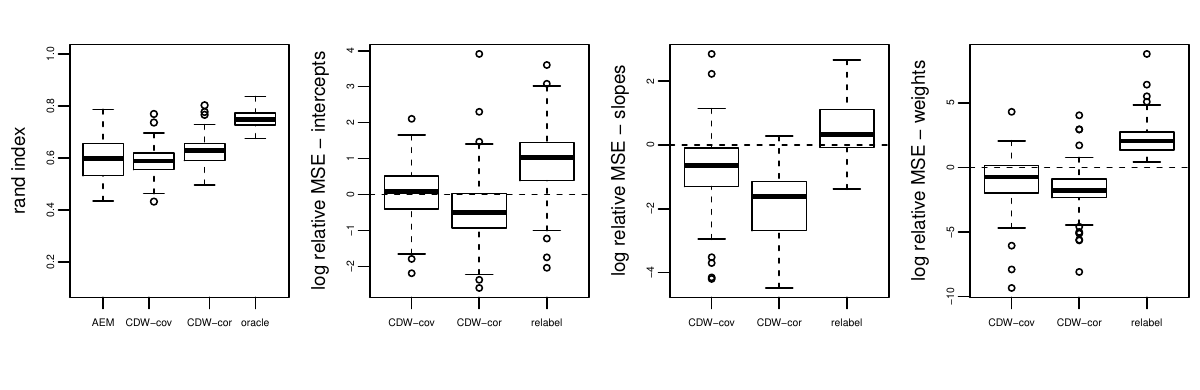}
	\captionof{figure}{\label{fig:B2summary} Results from setting B2: Rand index (left) and MSE relative to anchored EM for the intercept parameters (left center), slope parameters (right center), and mixture weights (right). }
\vskip
4mm
The estimated parameters in both settings are shown in Table~\ref{sB:estimates}. In all methods, although three distinct intercepts are
estimated, the average posterior means of the high- and low-intercept
lines are under- and over-estimated, respectively. This is expected
behavior in a mixture model because uncertainty in classifications
will lead to component means being shrunk towards each other.  
\vskip
4mm
	\captionof{table}{ \label{sB:estimates}Average posterior means for each model under setting B1 (top) and B2 (bottom). Values are the posterior means for each parameter, averaged over all data sets. Values in parentheses are estimated standard errors.}
	\begin{adjustbox}{width=\textwidth}
		\begin{tabular}{rllllllllll}
			\hline
			\hline
			\multicolumn{1}{l}{}&		\multicolumn{10}{l}{\textit{Setting B1}}\\
						\hline
			\hline
			\multicolumn{1}{l}{}&		\multicolumn{3}{|c|}{Component 1}&\multicolumn{3}{|c|}{Component 2}&\multicolumn{3}{|c|}{Component 3}&\multicolumn{1}{c}{}\\
			\hline
			& $\beta_{0}$ & $\beta_{1}$ & $\eta$ &  $\beta_{0}$ & $\beta_{1}$ & $\eta$ &  $\beta_{0}$ & $\beta_{1}$ & $\eta$ & $\sigma^2$ \\ 
			\hline
true &    -3 &   0.7 & 0.3333 &  -2.5 &   0.7 & 0.3333 &    -2 &   0.7 & 0.3333 & 0.008772 \\ 
AEM & -2.842 (0.08)  & 0.711 (0.10)  & 0.356 (0.07)  & -2.501 (0.13)  & 0.672 (0.15)  & 0.296 (0.05)  & -2.155 (0.08)  & 0.703 (0.09)  & 0.348 (0.07)  & 0.078 (0.01)  \\ 
CDW-cov & -2.809 (0.08)  & 0.695 (0.08)  & 0.359 (0.08)  & -2.496 (0.20)  & 0.692 (0.14)  & 0.289 (0.04)  & -2.194 (0.10)  & 0.701 (0.08)  & 0.351 (0.08)  & 0.085 (0.01)  \\ 
CDW-cor & -2.869 (0.05)  & 0.691 (0.05)  & 0.358 (0.05)  & -2.498 (0.10)  & 0.707 (0.07)  & 0.289 (0.03)  & -2.133 (0.05)  & 0.696 (0.05)  & 0.353 (0.05)  & 0.076 (0.01)  \\ 
relabel & -2.709 (0.11)  & 0.685 (0.05)  & 0.476 (0.15)  & -2.502 (0.03)  & 0.355 (0.05)  & 0.064 (0.02)  & -2.293 (0.11)  & 0.674 (0.07)  & 0.459 (0.16)  & 0.115 (0.02)  \\ 
			\hline
			\hline
			\multicolumn{1}{l}{}&		\multicolumn{10}{l}{\textit{Setting B2}}\\
			\hline
			\hline
			\multicolumn{1}{l}{}&		\multicolumn{3}{|c|}{Component 1}&\multicolumn{3}{|c|}{Component 2}&\multicolumn{3}{|c|}{Component 3}&\multicolumn{1}{c}{}\\
			\hline
			& $\beta_{0}$ & $\beta_{1}$ & $\eta$ &  $\beta_{0}$ & $\beta_{1}$ & $\eta$ &  $\beta_{0}$ & $\beta_{1}$ & $\eta$ & $\sigma^2$ \\ 
						\hline
true &    -3 &   0.7 & 0.3333 &  -2.5 &   0.7 & 0.3333 &    -2 &   0.7 & 0.3333 & 0.04167 \\ 
AEM & -2.873 (0.09)  & 0.697 (0.13)  & 0.310 (0.08)  & -2.529 (0.16)  & 0.657 (0.20)  & 0.346 (0.07)  & -2.155 (0.08)  & 0.688 (0.13)  & 0.344 (0.09)  & 0.110 (0.01)  \\ 
CDW-cov & -2.818 (0.16)  & 0.671 (0.10)  & 0.335 (0.07)  & -2.485 (0.19)  & 0.711 (0.15)  & 0.325 (0.05)  & -2.218 (0.15)  & 0.674 (0.12)  & 0.340 (0.07)  & 0.115 (0.02)  \\ 
CDW-cor & -2.866 (0.10)  & 0.692 (0.07)  & 0.331 (0.04)  & -2.517 (0.13)  & 0.671 (0.10)  & 0.317 (0.04)  & -2.156 (0.12)  & 0.691 (0.07)  & 0.352 (0.05)  & 0.108 (0.01)  \\ 
relabel & -2.662 (0.11)  & 0.660 (0.07)  & 0.449 (0.19)  & -2.517 (0.05)  & 0.367 (0.07)  & 0.074 (0.06)  & -2.352 (0.10)  & 0.671 (0.07)  & 0.477 (0.19)  & 0.158 (0.02)  \\ 
			\hline
		\end{tabular}
		
	\end{adjustbox}

\clearpage
\subsection{Results: Setting~C}
Figure~\ref{modelsC} shows the regression lines and sample data sets under Settings C1 and C2. The model is characterized by two parallel lines with a third line intersecting both. The points generated by setting C2 are particularly difficult to distinguish as being generated from different groups. 
\vskip 4mm
	\includegraphics[width=\textwidth]{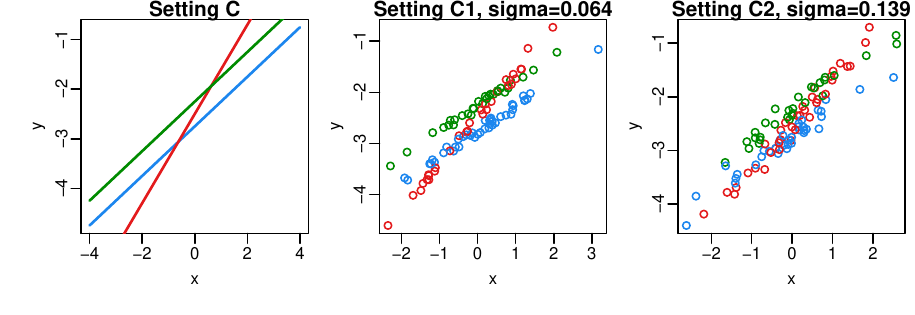}
        \captionof{figure}{\label{modelsC}
True regression lines and
    representative simulated data sets for Setting~C. Plotting symbols are
  colored by their true allocation.}

\hspace*{1.7pc}  The estimation accuracy of the four methods is shown in the right panels of Figures~\ref{fig:C1summary} and~\ref{fig:C2summary} for settings~C1 and~C2, respectively. In setting C1, the anchored EM has the strongest accuracy in parameter estimation, while the CDW-cor method has the highest clustering accuracy. 
	\includegraphics[width=.98\textwidth]{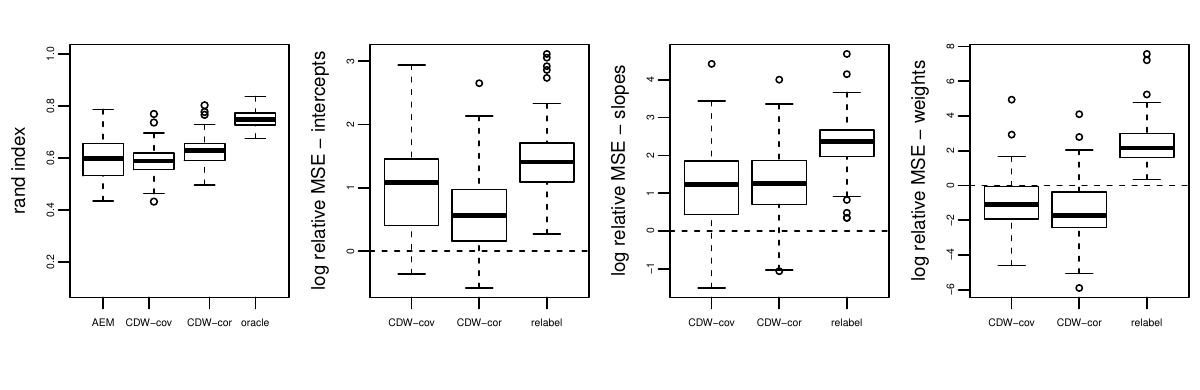}
\captionof{figure}{\label{fig:C1summary} Results from setting C1: Rand index (left) and MSE relative to anchored EM for the intercept parameters (left center), slope parameters (right center), and mixture weights (right).} 

	\includegraphics[width=.98\textwidth]{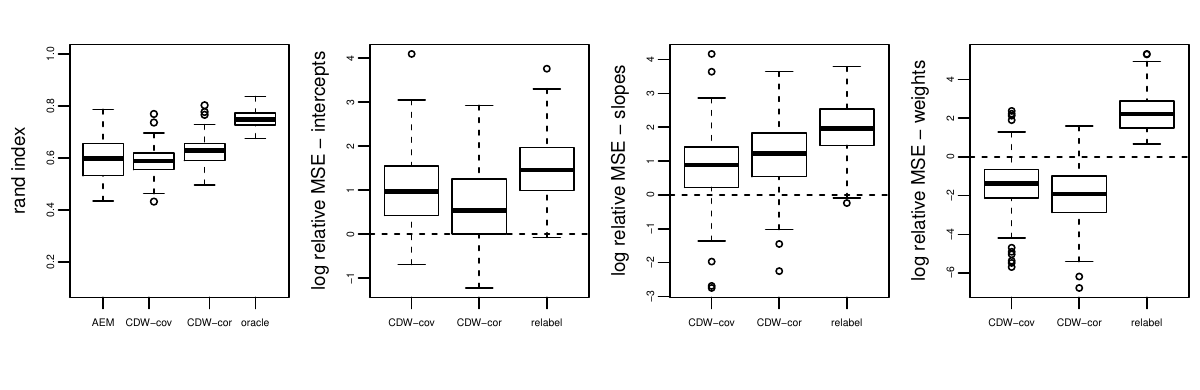}
\captionof{figure}{\label{fig:C2summary} Results from setting C2: Rand index (left) and MSE relative to anchored EM for the intercept parameters (left center), slope parameters (right center), and mixture weights (right).}

A very similar pattern is seen in setting C2, as shown in
Figure~\ref{fig:C2summary}. The estimates in Table~\ref{sC:estimates}
indicate that there is particularly high error in estimating the steep
slope of the intersecting line under the two CDW methods. In addition, both CDW
methods have average intercepts that are closer together than the true
values. The anchored EM also results in some shrinkage of the
intercept estimates towards each other, but to a lesser degree.
\vskip 4mm
\captionof{table}{\label{sC:estimates}Average posterior means for each model under setting C1 (top) and C2 (bottom). Values are the posterior means for each parameter, averaged over all data sets. Values in parentheses are estimated standard errors.} 
	\begin{adjustbox}{width=\textwidth}
		\begin{tabular}{rllllllllll}
			\hline
			\hline
			\multicolumn{1}{l}{}&		\multicolumn{10}{l}{\textit{Setting C1}}\\
			\hline
			\hline
			\multicolumn{1}{l}{}&		\multicolumn{3}{|c|}{Component 1}&\multicolumn{3}{|c|}{Component 2}&\multicolumn{3}{|c|}{Component 3}&\multicolumn{1}{c}{}\\
			\hline
			& $\beta_{0}$ & $\beta_{1}$ & $\eta$ &  $\beta_{0}$ & $\beta_{1}$ & $\eta$ &  $\beta_{0}$ & $\beta_{1}$ & $\eta$ & $\sigma^2$ \\ 
			\hline
true & -2.75 &   0.5 & 0.3333 &  -2.5 &   0.9 & 0.3333 & -2.25 &   0.5 & 0.3333 & 0.004064 \\ 
AEM & -2.633 (0.04)  & 0.494 (0.06)  & 0.281 (0.05)  & -2.500 (0.03)  & 0.825 (0.07)  & 0.423 (0.09)  & -2.360 (0.04)  & 0.491 (0.06)  & 0.297 (0.06)  & 0.054 (0.00)  \\ 
CDW-cov & -2.561 (0.06)  & 0.614 (0.16)  & 0.335 (0.06)  & -2.495 (0.05)  & 0.640 (0.19)  & 0.330 (0.07)  & -2.434 (0.06)  & 0.618 (0.14)  & 0.336 (0.05)  & 0.060 (0.01)  \\ 
CDW-cor & -2.597 (0.06)  & 0.574 (0.12)  & 0.315 (0.04)  & -2.498 (0.05)  & 0.693 (0.15)  & 0.358 (0.05)  & -2.401 (0.05)  & 0.592 (0.13)  & 0.327 (0.04)  & 0.059 (0.01)  \\ 
relabel & -2.515 (0.03)  & 0.559 (0.21)  & 0.404 (0.22)  & -2.499 (0.03)  & 0.388 (0.22)  & 0.237 (0.26)  & -2.476 (0.03)  & 0.510 (0.19)  & 0.360 (0.20)  & 0.069 (0.01)  \\  
			\hline
			\hline
			\multicolumn{1}{l}{}&		\multicolumn{10}{l}{\textit{Setting C2}}\\
			\hline
			\hline
			\multicolumn{1}{l}{}&		\multicolumn{3}{|c|}{Component 1}&\multicolumn{3}{|c|}{Component 2}&\multicolumn{3}{|c|}{Component 3}&\multicolumn{1}{c}{}\\
			\hline
			& $\beta_{0}$ & $\beta_{1}$ & $\eta$ &  $\beta_{0}$ & $\beta_{1}$ & $\eta$ &  $\beta_{0}$ & $\beta_{1}$ & $\eta$ & $\sigma^2$ \\ 
			\hline
true & -2.75 &   0.5 & 0.3333 &  -2.5 &   0.9 & 0.3333 & -2.25 &   0.5 & 0.3333 & 0.01931 \\ 
AEM & -2.642 (0.05)  & 0.503 (0.13)  & 0.285 (0.07)  & -2.503 (0.04)  & 0.793 (0.13)  & 0.404 (0.10)  & -2.363 (0.05)  & 0.512 (0.10)  & 0.312 (0.07)  & 0.068 (0.01)  \\ 
CDW-cov & -2.567 (0.06)  & 0.623 (0.15)  & 0.343 (0.05)  & -2.502 (0.06)  & 0.609 (0.20)  & 0.325 (0.08)  & -2.430 (0.06)  & 0.620 (0.17)  & 0.332 (0.06)  & 0.075 (0.01)  \\ 
CDW-cor & -2.618 (0.05)  & 0.606 (0.12)  & 0.318 (0.04)  & -2.496 (0.05)  & 0.662 (0.13)  & 0.356 (0.04)  & -2.392 (0.06)  & 0.606 (0.13)  & 0.326 (0.04)  & 0.074 (0.01)  \\ 
relabel & -2.527 (0.03)  & 0.535 (0.20)  & 0.382 (0.23)  & -2.502 (0.03)  & 0.384 (0.18)  & 0.201 (0.24)  & -2.478 (0.04)  & 0.548 (0.20)  & 0.417 (0.24)  & 0.084 (0.01)  \\
			\hline
		\end{tabular}
		
	\end{adjustbox}
%	\caption{ \label{sC:estimates}Average posterior means for each model under setting C1 (top) and C2 (bottom). Values are the posterior means for each parameter, averaged over all data sets. Values in parentheses are estimated standard errors.} 
%\end{table}

\clearpage
	\subsection{Setting D: three predictors} \label{section:simD}
	In addition, we considered one multiple regression case in which three numeric predictors were used. We generated three numeric predictors, $x_1$, $x_2$, $x_3$ with the following correlation matrix:
	\begin{eqnarray}
	\begin{bmatrix}
1 & 0.8 & 0.05 \\
0.8 & 1 & -0.10 \\
0.05 & -0.10 & 1
\end{bmatrix}
\end{eqnarray}

The regression coefficients for each component were:
\begin{align*}
	 \vv{\beta}_1 = (-3.0,  0.7, -1.6,  0.2)^\prime; \;\; \vv{\beta}_2 = (-2.5,  0.9, -1.6, -0.2)^\prime; \vv{\beta}_3 = (-2.0,  0.5, -1.6,  0.0)^\prime.
	 \end{align*}
The residual standard deviation, $\sigma$, was set to be
0.224.
	  
\hspace*{1.7pc} Figure~\ref{fig:Dsummary} summarizes the
performance of each of the methods. For this setting, the EM and CDW
methods tend to have similar MSEs for the coefficients corresponding
to the numeric predictors. The error is somewhat higher for CDW-cov in
estimating the intercept. As in the other settings, we see the highest
accuracy in clustering from CDW-cor and the lowest from CDW-cov. The
parameter estimates are given in Table~\ref{sD:estimates}. All methods
typically produce accurate estimates of $\beta_2$, which does not
differ across the components. The $\beta_1$ coefficient, associated
with the predictor $x_1$ which is highly correlated with
$x_2$, is also estimated with accuracy. The
intercepts, as in the simpler models, tend to shrink together with the
CDW-cov method exhibiting this behavior to the greatest degree.
	
	\includegraphics[width=.98\textwidth]{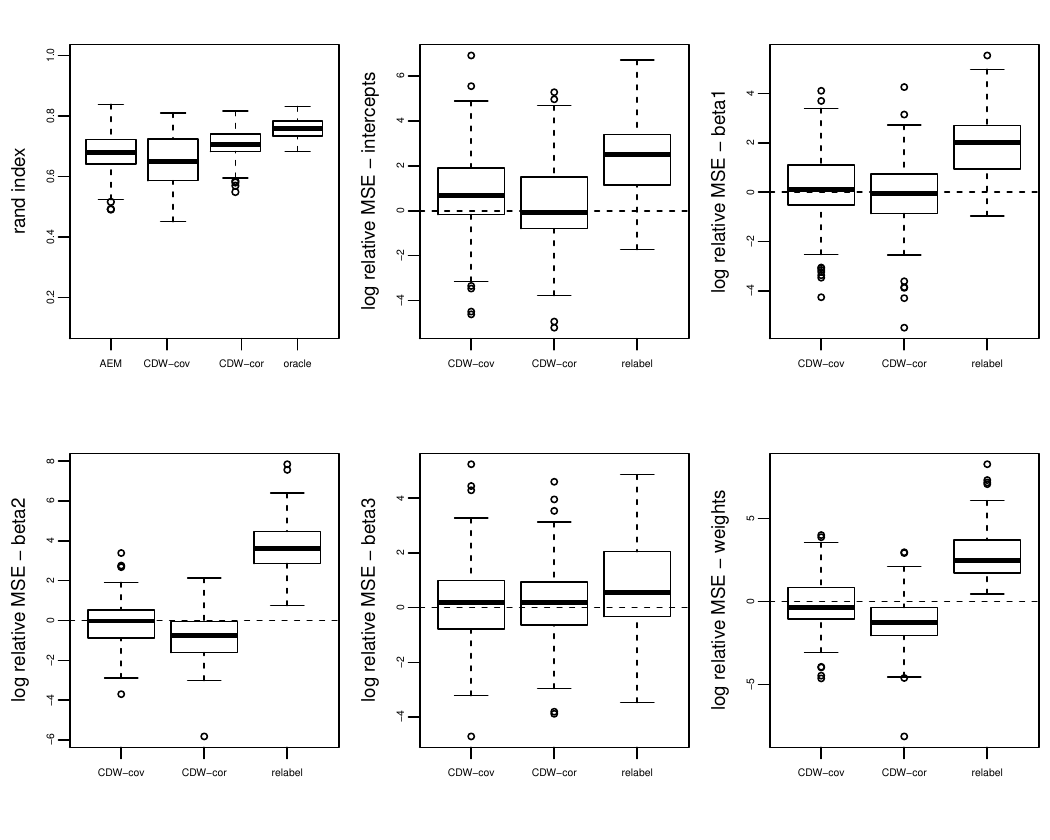}	\captionof{figure}{\label{fig:Dsummary} Results from setting D: Rand index (top left) and MSE relative to anchored EM for the regression coefficients and mixture weights. }
\clearpage

%\begin{table}[ht]
\captionof{table}{\label{sD:estimates}Average posterior means for each model under setting D. Values are the posterior means for each parameter, averaged over all data sets. Values in parentheses are estimated standard errors.} 
		\begin{adjustbox}{width=.94\textwidth}
	\begin{tabular}{rllllllllllllllll}
		\hline
				\hline
		\multicolumn{1}{l}{}&		\multicolumn{16}{l}{\textit{Setting D}}\\
		\hline
		\hline
		\multicolumn{1}{l}{}&		\multicolumn{5}{|c|}{Component 1}&\multicolumn{5}{|c|}{Component 2}&\multicolumn{5}{|c|}{Component 3}&\multicolumn{1}{c}{}\\
		\hline
		& $\beta_0$ & $\beta_1$ & $\beta_2$ &  $\beta_3$ & $\eta$ & $\beta_0$ & $\beta_1$ & $\beta_2$ &  $\beta_3$ & $\eta$ & $\beta_0$ & $\beta_1$ & $\beta_2$ &  $\beta_3$ &$\eta$ & $\sigma^2$ \\ 
  \hline
true &    -3 &   0.7 &  -1.6 &   0.2 & 0.3333 &  -2.5 &   0.9 &  -1.6 &  -0.2 & 0.3333 &    -2 &   0.5 &  -1.6 &     0 & 0.3333 &  0.05 \\ 
AEM & -2.86 (0.19)  & 0.68 (0.18)  & -1.56 (0.18)  & 0.24 (0.10)  & 0.32 (0.10)  & -2.47 (0.19)  & 0.80 (0.22)  & -1.59 (0.17)  & -0.14 (0.14)  & 0.38 (0.10)  & -2.18 (0.24)  & 0.46 (0.24)  & -1.55 (0.21)  & 0.02 (0.14)  & 0.30 (0.09)  & 0.09 (0.02)  \\ 
CDW-cov & -2.76 (0.24)  & 0.68 (0.21)  & -1.60 (0.19)  & 0.21 (0.11)  & 0.30 (0.08)  & -2.47 (0.19)  & 0.77 (0.18)  & -1.60 (0.14)  & -0.12 (0.16)  & 0.38 (0.09)  & -2.27 (0.24)  & 0.52 (0.24)  & -1.56 (0.20)  & 0.00 (0.12)  & 0.32 (0.08)  & 0.11 (0.02)  \\ 
CDW-cor & -2.83 (0.23)  & 0.69 (0.17)  & -1.58 (0.14)  & 0.17 (0.12)  & 0.33 (0.05)  & -2.45 (0.23)  & 0.76 (0.17)  & -1.59 (0.11)  & -0.10 (0.12)  & 0.35 (0.05)  & -2.22 (0.27)  & 0.57 (0.18)  & -1.59 (0.12)  & -0.01 (0.09)  & 0.32 (0.05)  & 0.10 (0.02)  \\ 
relabel & -2.60 (0.25)  & 0.45 (0.29)  & -1.17 (0.50)  & 0.12 (0.10)  & 0.33 (0.21)  & -2.42 (0.23)  & 0.63 (0.21)  & -1.52 (0.24)  & -0.04 (0.11)  & 0.60 (0.21)  & -2.50 (0.07)  & 0.06 (0.16)  & -0.24 (0.36)  & 0.02 (0.04)  & 0.06 (0.16)  & 0.17 (0.05)  \\ 
\hline
	\end{tabular}
	\end{adjustbox}

%	\caption{ \label{sD:estimates}Average posterior means for each model under setting D. Values are the posterior means for each parameter, averaged over all data sets. Values in parentheses are estimated standard errors.} 

%\end{table}
	
	\section{Sensitivity analysis} \label{supplement:sensitivity}
	In this section, we assess the sensitivity of our results to the number of mixture components, the strength of prior assumptions, and to the number of anchor points. 
	
	\subsection{Number of mixture components}
	Here, we show the results of fitting a mixture with $k=2$ components to the mammals data. The main text discussed that in selecting the number of components, two or three components are preferred.  Figure~\ref{fig:k2} shows the anchor points and estimated regression lines resulting from the two-component model. 
	
		\includegraphics[scale=.9]{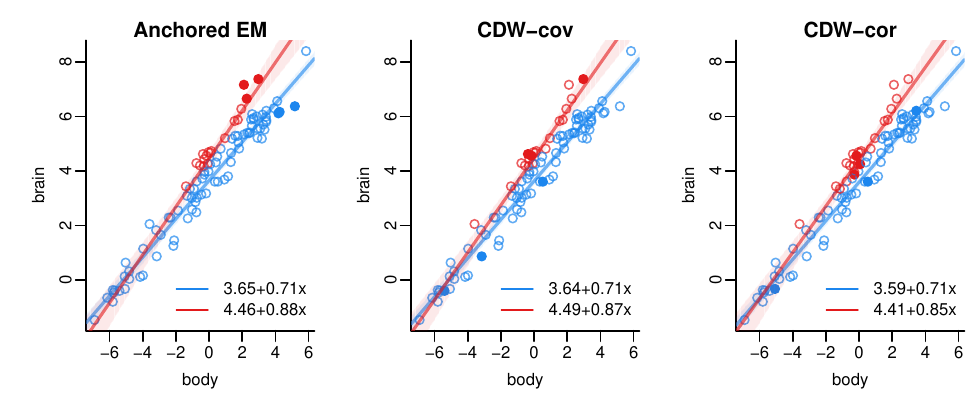}\captionof{figure}{\label{fig:k2} Anchor points and estimated regression lines for a model with two components. }
		Compared to the 3-component model presented in the main text, the estimated red component captures the species that constitute a group with a higher slope than the others, including some of the large primates. This sub-group is similar to Component~3 identified in the main text, although, in the two-component fit, the estimated regression line is somewhat less steep and has a smaller intercept than Component~3 in the main text. The second group, shown in blue in Figure~\ref{fig:k2}, has an estimated slope of 0.71 for all anchor models, which is similar to the slopes of Components~1 and~2 in the main text. The estimated intercepts of this blue group range from 4.41 to 4.49, which falls between the intercepts estimated for Components~1 and ~2 in the main text. The similarities between the lines estimated in the $k=2$ and $k=3$ models indicate that both models are able to partition the species into similar subgroups, and that the 3-component model further refines the grouping by capturing differences in average brain size. 
	\subsection*{Sensitivity to hyperparameters}
	The main text presents results 
	under the following prior specification:
	\begin{align} \label{prior}
	\beta_j \sim N_2((3.5,0.6)^\prime, \begin{bmatrix}
	1 &0 \\0 & 0.5
	\end{bmatrix}) \\
	\sigma^{-2} \sim \text{Gamma}(\text{shape=5},\text{rate=1}) 
	\end{align}
We will assess the effect of changing these hyperparameters on conclusions from the analysis of the mammals data.  Table~\ref{table:mammalsci} gives the posterior means and 90\% credible intervals under the original hyperparameters for reference.

\hspace*{1.7pc}  The anchored EM method includes update steps that
depend on the prior hyperparameters. The CDW methods select anchor
points based on a preliminary simple linear regression fit, which may
also be affected by specification of the hyperparameters. Because of
this, in assessing sensitivity to the hyperparameters, we do not hold
the anchor points fixed, but re-select them for each case.

\clearpage

\captionof{table}{\label{table:mammalsci} Posterior means and (90\% credible intervals) of the regression coefficients for the mammals data. Estimates are conditional on the hyperparameters used in the main text.}
\begin{adjustbox}{width=\textwidth}
	\begin{tabular}{rllllll}
		\hline
		\multicolumn{1}{l}{}&		\multicolumn{2}{c}{Component 1}&\multicolumn{2}{c}{Component 2}&\multicolumn{2}{c}{Component 3}\\
\hline
& 
$\beta_0$ & $\beta_1$ & $\beta_0$ & $\beta_1$ & $\beta_0$ & $\beta_1$ \\ 
		\hline
		AEM & 3.39 (3.18, 3.60) & 0.697 (0.65, 0.74) & 3.97 (3.72, 4.24) & 0.712 (0.64, 0.77) & 4.56 (4.35, 4.76) & 0.911 (0.83, 1.02) \\ 
		CDW-cov & 3.43 (3.22, 3.65) & 0.695 (0.65, 0.74) & 4.00 (3.75, 4.26) & 0.691 (0.59, 0.76) & 4.53 (4.31, 4.73) & 0.915 (0.83, 1.02) \\ 
		CDW-cor & 3.47 (3.21, 3.83) & 0.694 (0.61, 0.75) & 3.83 (3.54, 4.08) & 0.724 (0.67, 0.78) & 4.52 (4.32, 4.70) & 0.891 (0.81, 0.99) \\ 
		\hline
	\end{tabular}
\end{adjustbox}
\vskip 8mm
\paragraph{Sensitivity to $V$.} We first consider the sensitivity of
the results to the strength of prior information on $\vv{\beta}$ by
considering prior variances of $4$ and $2$ on the intercepts and
slopes, respectively. These variances are larger than those specified
in the main text
analyses by a factor of $4$. The prior means were left unchanged.

\hspace*{1.7pc} Table~\ref{table:v0} gives the posterior means and
credible intervals for the regression coefficients. The estimates are
very similar to those in Table~\ref{table:mammalsci} under the A-EM
and CDW-cor method. For the CDW-cov method, the estimates for the
steepest regression line (labeled Component~3) are similar to their
original values. For the two shallower lines (Component~1 and~2), the
posterior credible interval of the
estimated intercepts overlap to a much greater degree
under this less-informative prior.
\vskip 1mm
\captionof{table}{\label{table:v0} Posterior means and (90\% credible intervals) of the regression coefficients with $v_0=4$, $v_1=2$.}
\begin{adjustbox}{width=\textwidth}
	\begin{tabular}{rllllll}
		\hline 
		\multicolumn{1}{l}{}&		\multicolumn{2}{c}{Component 1}&\multicolumn{2}{c}{Component 2}&\multicolumn{2}{c}{Component 3}\\
		\hline
		& 
		$\beta_0$ & $\beta_1$ & $\beta_0$ & $\beta_1$ & $\beta_0$ & $\beta_1$ \\ 
		\hline
		AEM & 3.39 (3.16, 3.60) & 0.699 (0.66, 0.74) & 3.98 (3.73, 4.25) & 0.712 (0.64, 0.77) & 4.58 (4.37, 4.78) & 0.910 (0.82, 1.02) \\ 
		CDW-cov & 3.55 (3.20, 3.96) & 0.732 (0.67, 0.81) & 3.79 (3.52, 4.07) & 0.674 (0.58, 0.74) & 4.56 (4.32, 4.77) & 0.895 (0.79, 1.03) \\ 
		CDW-cor & 3.43 (3.19, 3.75) & 0.696 (0.63, 0.75) & 3.87 (3.61, 4.09) & 0.726 (0.68, 0.78) & 4.55 (4.36, 4.73) & 0.897 (0.81, 1.00) \\ 
		\hline
	\end{tabular}
\end{adjustbox}
\vskip 8mm
\paragraph{Sensitivity to $a, b$.} We next consider the sensitivity of results to the strength of prior information on $\sigma^{-2}$ by setting the gamma shape and rate to be $a=0.5$ and $b=0.1$, respectively.  This specification leaves the prior mean of $\sigma^{-2}$ unchanged, but increases the prior variance from 5 to 50. Table~\ref{table:ab} shows the estimated posterior means under this weaker prior. The slope of Component~2 estimated by CDW-cor is smaller than its estimate under the original hyperparameters and the slope of Component~1 is larger.  The credible intervals of these parameters nonetheless overlap with the original estimates.

\captionof{table}{\label{table:ab} Posterior means and (90\% credible intervals) of the regression coefficients with $a=0.5$, $b=0.1$.}
\begin{adjustbox}{width=\textwidth}
	\begin{tabular}{rllllll}
		\hline 
		\multicolumn{1}{l}{}&		\multicolumn{2}{c}{Component 1}&\multicolumn{2}{c}{Component 2}&\multicolumn{2}{c}{Component 3}\\
		\hline
		& 
		$\beta_0$ & $\beta_1$ & $\beta_0$ & $\beta_1$ & $\beta_0$ & $\beta_1$ \\  
		\hline
		AEM & 3.35 (3.17, 3.55) & 0.701 (0.66, 0.74) & 3.98 (3.76, 4.21) & 0.709 (0.64, 0.77) & 4.58 (4.40, 4.75) & 0.903 (0.83, 0.99) \\ 
		CDW-cov & 3.40 (3.15, 3.95) & 0.687 (0.54, 0.74) & 3.87 (3.61, 4.06) & 0.723 (0.67, 0.77) &   4.58 (4.40, 4.75) & 0.889 (0.82, 0.97) \\ 
		CDW-cor &  3.48 (3.19, 3.78) & 0.735 (0.67, 0.81) &3.82 (3.44, 4.23) & 0.648 (0.49, 0.74) & 4.50 (4.29, 4.68) & 0.865 (0.79, 0.93) \\ 
		\hline
	\end{tabular}
\end{adjustbox}
\vskip 4mm

\subsection{Sensitivity to $m$} \label{supplement:msensitivity} While
selection of the number of anchor points is not a focus of our study,
we considered the estimates resulting from using the proposed
anchoring procedures with fewer anchor points.  Intuitively, at least
two anchor points per component should be needed to pin down the
component-specific linear regression. Specification of a single anchor
point per component, while sufficient to avoid label switching, may
not be enough to provide accurate modeling.  The table below displays
posterior means for the three anchoring methods with 1 and 2 points
per component.

\hspace*{1.7pc} The top panel of Table~\ref{table:m} shows the
estimated regression coefficients under the anchor model with one
point per component. Compared to the fit with $m=3$, the credible
intervals are much wider, which is a natural consequence of a model
with weaker prior information. In these models, all methods have
identified a line with a steep slope and large intercept, arbitrarily
labeled Component~3, although the estimated slopes for this line are
slightly smaller in the CDW models than their original fits. For all
methods, the credible intervals for the regression parameters of
Components~1 and~2 overlap substantially, particularly those estimated
by the CDW-cov and A-EM methods.

\hspace*{1.7pc} With two anchor points per component, the results
summarized in the bottom panel of Table~\ref{table:m} show that there
is a clearer separation of Component~1 and~2, with the former having a
small intercept with credible intervals that do not overlap with those
of the intercepts for the other components under the A-EM and CDW-cor
methods. As in the $m=1$ case, the estimates for Component~3 are
similar to those obtained with the original analysis.

\captionof{table}{\label{table:m} Posterior means and (90\% credible intervals) of the regression coefficients with $m=1$ (top) and $m=2$ (bottom). }
				\begin{adjustbox}{width=\textwidth}
	\begin{tabular}{rllllll}
		\hline
		\multicolumn{7}{c}{$m=1$}\\
				\hline 
		\multicolumn{1}{l}{}&		\multicolumn{2}{c}{Component 1}&\multicolumn{2}{c}{Component 2}&\multicolumn{2}{c}{Component 3}\\
		\hline
		& 
		$\beta_0$ & $\beta_1$ & $\beta_0$ & $\beta_1$ & $\beta_0$ & $\beta_1$ \\ 
		\hline
		AEM & 3.68 (3.22, 4.43) & 0.635 (0.41, 0.73) & 3.84 (3.52, 4.27) & 0.725 (0.66, 0.80) & 4.50 (4.22, 4.76) & 0.917 (0.80, 1.10) \\ 
		CDW-cov & 3.75 (3.22, 4.53) & 0.773 (0.67, 1.01) & 3.77 (3.40, 4.22) & 0.679 (0.52, 0.75) & 4.46 (4.00, 4.75) & 0.835 (0.52, 1.00) \\ 
		CDW-cor & 3.56 (3.24, 3.99) & 0.711 (0.63, 0.79) & 3.99 (3.49, 4.65) & 0.739 (0.61, 0.93) & 4.40 (3.83, 4.70) & 0.852 (0.62, 1.02) \\ 
		\hline
				\multicolumn{7}{c}{$m=2$}\\
		  \hline
		AEM & 3.42 (3.17, 3.66) & 0.697 (0.65, 0.74) & 3.97 (3.69, 4.28) & 0.714 (0.64, 0.78) & 4.55 (4.31, 4.77) & 0.914 (0.82, 1.05) \\ 
		CDW-cov & 3.58 (3.20, 4.22) & 0.660 (0.46, 0.75) & 3.80 (3.53, 4.08) & 0.724 (0.67, 0.77) & 4.53 (4.29, 4.75) & 0.899 (0.80, 1.02) \\ 
		CDW-cor & 3.48 (3.24, 3.70) & 0.711 (0.67, 0.76) & 4.05 (3.61, 4.61) & 0.735 (0.58, 0.93) & 4.37 (3.91, 4.64) & 0.820 (0.63, 0.94) \\ 
		\hline
%						\multicolumn{7}{c}{Three}\\
%		  \hline
%		AEM & 3.39 (3.17, 3.60) & 0.698 (0.66, 0.74) & 3.97 (3.72, 4.23) & 0.712 (0.64, 0.77) & 4.56 (4.35, 4.76) & 0.910 (0.82, 1.02) \\ 
%		CDW-cov & 3.66 (3.27, 4.24) & 0.640 (0.46, 0.73) & 3.72 (3.49, 3.97) & 0.739 (0.69, 0.80) & 4.52 (4.30, 4.71) & 0.891 (0.80, 0.99) \\ 
%		CDW-cor & 3.45 (3.23, 3.66) & 0.707 (0.66, 0.76) & 3.95 (3.56, 4.49) & 0.747 (0.65, 0.88) & 4.33 (3.91, 4.60) & 0.778 (0.64, 0.87) \\ 
%		\hline
%								\multicolumn{7}{c}{Four}\\
%		  \hline
%		& beta0 & beta1 & beta0.1 & beta1.1 & beta0.2 & beta1.2 \\ 
%		\hline
%		AEM & 3.36 (3.16, 3.55) & 0.699 (0.66, 0.74) & 3.97 (3.76, 4.20) & 0.708 (0.64, 0.76) & 4.57 (4.38, 4.75) & 0.907 (0.83, 1.00) \\ 
%		CDW-cov & 3.34 (3.16, 3.56) & 0.713 (0.67, 0.76) & 3.93 (3.74, 4.12) & 0.712 (0.66, 0.76) & 4.60 (4.42, 4.78) & 0.893 (0.81, 0.99) \\ 
%		CDW-cor & 3.38 (3.20, 3.59) & 0.705 (0.67, 0.75) & 3.89 (3.67, 4.11) & 0.689 (0.61, 0.75) & 4.50 (4.33, 4.67) & 0.858 (0.78, 0.94) \\ 
%		\hline
	\end{tabular}
\end{adjustbox}

	\section{Model-based clustering and known taxonomy} \label{supplement:clustering}
	
\begin{figure}[t!]
	\includegraphics[width=.9\textwidth]{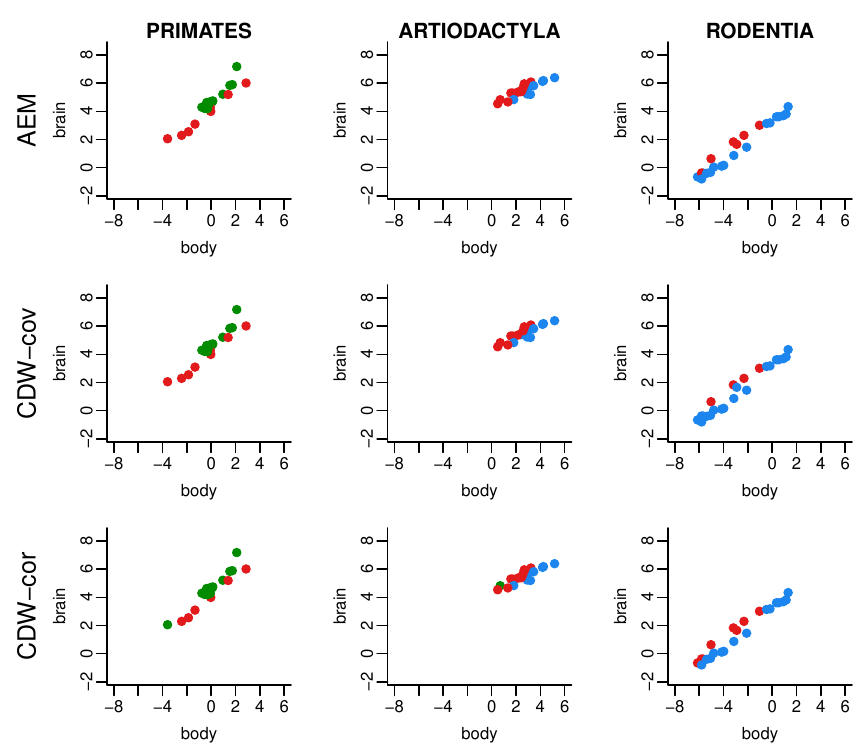}
	\caption{\label{figure:clusters_byorder} Data for the species
		in orders Primates, Artiodactyla, and Rodentia color-coded
		according to their model-based $\widehat{\vv{s}}$.  The rows
		correspond to the three anchor models.  The color
                coding used to distinguish the three mixture
                components matches that 
                used in Figure~4 of the main article.} 
\end{figure}

The estimated component assignments $\widehat{\vv{s}}$ give a
model-based grouping of the species which ignores the additional data
on the species' taxonomic orders and suborders. A comparison of these
groups with the true taxonomy of the species can shed light on the
allometric questions posed at the beginning of this article: the
species assigned to the same component by $\widehat{\vv{s}}$ have
similar estimated regression slopes, and if there is a correspondence
between $\widehat{\vv{s}}$ and the species' true orders, it may
indicate that certain taxonomic groups have distinct body mass/brain
mass relationships.  Figure~\ref{figure:clusters_byorder} displays the
data from species in three of the 13 orders, color-coded according to the
model-based cluster assignment.  The displayed orders are Primates,
Artiodactyla, and Rodentia, which account for 21, 21, and 24 of the
100 species in the data, respectively.  

\hspace*{1.7pc} The first column of the figure shows the Primate
species, which, for all three models, are split between Components~2
and~3.  For a given body mass, most larger-brained species are
assigned to Component~3 and most of those with smaller brains are
assigned to Component~2.  Component~3 also contains the three species
of the Cetacea order (not pictured) under all three model fits.  Two
primate species are assigned to Component~3 by the CDW-cor method and
to Component~2 by the other two models: Hapale Leucocephala and
Macaca Maurus, both of the sub-order Anthropoidea.  The Hapale
Leucocephala is the primate with the smallest body, but this species'
brain is actually large given its body mass. Component~2 also contains
all three primate species of the Prosimii sub-order. So, the mixture
model is sensitive to this aspect of the taxonomic classification,
recognizing that the Prosimii species have small brains given their
body masses.

\hspace*{1.7pc} The clustering among the Rodentia differs most across
the three anchor models. The CDW-cov model assigns only four of 24
Rodentia to Component~2, while the CDW-cor method assigns 9. All of
the largest-bodied Rodentia species, seen as the far right points
colored in blue, are assigned to Component~1 by all of the
models. Interestingly, one species that falls next to these points in
the plot, the Myocastor Coypus, is assigned to Component~2 by all
models, in spite of its similar body size to the neighboring
points. This indicates a sensitivity of all models to its slight
decrease in brain size compared to the adjacent species, whose body
sizes are very similar.
% A partial explanation for
%this is that larger body masses are a characterizing feature of these
%species, but the EM-reg method chooses anchor points
%conditional on $x$ and thus is less likely to identify groups that are
%characterized by differences in $x$-values.
	\section*{Acknowledgments}
	
	This material is based on work supported by the National Science
	Foundation under grant no.~SES-1424481 and SES-1921523.
	\par
	
	%%%%%%%%%%%%%%%%%%%%%%%%%%%%%%%%%%%%%%%%%%%%%%%%%%%%%%%%%%%%%%%%%%%%%%%%%%%%%%%%%%%%%%%%%%

	\renewcommand{\baselinestretch}{1.5}

%\iffalse
\bibhang=1.7pc
\bibsep=2pt
\fontsize{9}{14pt plus.8pt minus .6pt}\selectfont
\renewcommand\bibname{\large \bf References}
%\begin{thebibliography}{11}
\expandafter\ifx\csname
natexlab\endcsname\relax\def\natexlab#1{#1}\fi
\expandafter\ifx\csname url\endcsname\relax
  \def\url#1{\texttt{#1}}\fi
\expandafter\ifx\csname urlprefix\endcsname\relax\def\urlprefix{URL}\fi
%\fi

	\renewcommand{\baselinestretch}{1}
\bibliographystyle{chicago}      % Chicago style, author-year citations
\bibliography{mybibfile_MP}   % name your BibTeX data base

%-------------------------------------------
%\vskip .65cm
%\noindent
%first author affiliation
%\vskip 2pt
%\noindent
%E-mail: dekunke@clemson.edu
%\vskip 2pt
%
%\noindent
%second author affiliation
%\vskip 2pt
%\noindent
%E-mail: (second author email)